\documentclass[journal]{IEEEtran}
\usepackage{graphicx}
\usepackage{amsmath,amssymb}
\usepackage{cite}
\usepackage{url}
\usepackage{booktabs}
\usepackage{float}
\usepackage{adjustbox}

\usepackage{etoolbox}

\begin{document}

\title{DP-JMRNet: A Deep Unfolding Network for Differential Phase Preservation in Sparse Bitemporal SAR Reconstruction}

\author{
Juncheng Bao, Zhen Zhang, and George P. Petropoulos%

\thanks{\emph{(Corresponding author: Zhen Zhang.)}}%
\thanks{Juncheng Bao is with the College of Information Science and Engineering, Hohai University, Changzhou 213200, China (e-mail: jcbao@hhu.edu.cn).}%
\thanks{Zhen Zhang is with the College of Information Science and Engineering, Hohai University, Changzhou 213200, China (e-mail: 20140005@hhu.edu.cn).}%
\thanks{George P. Petropoulos is with the Department of Geography, Harokopio University of Athens, 17671 Athens, Greece (e-mail: gpetropoulos@hua.gr).}%
}

\maketitle

\begin{abstract}
Complex SAR imagery is usually visualized and evaluated mainly through its magnitude. Phase is retained in the complex data but is rarely treated as a direct image-quality objective. Existing sparse reconstruction methods typically focus on magnitude fidelity and single-epoch complex reconstruction accuracy.  
However, the phase difference between two acquisitions is what drives
line-of-sight deformation retrieval in InSAR, from ground subsidence
monitoring to earthquake deformation mapping.
This paper proposes the Differential-Phase-Oriented Joint Masked
Reconstruction Network (DP-JMRNet), which uses deep unfolding to reconstruct the two epochs jointly
from masked observations under a differential-phase objective. An
exchange-equivariant interaction module makes the reconstruction independent
of epoch ordering. A coherence-aware gate opens cross-epoch sharing in
coherent regions and closes it where the two epochs disagree. On simulated bitemporal SAR data, DP-JMRNet attains the lowest differential-phase RMSE at 30\%, 40\%, and 50\% sampling rate, while maintaining competitive amplitude and complex-image fidelity.
This corresponds to a 47.5\% - 51.3\% reduction over the best baseline, achieved with
one third of its parameters. The same trend is validated on three Sentinel-1
scenes. A systematic study of acquisition design further
shows that sharing the same aperture support across epochs is necessary for
phase fidelity, whereas optimizing the  sampling mask does not improve
the differential phase. The code and data are available at
\url{https://github.com/JasonBao05/coherent-sar-unfolding}.

\end{abstract}


\newpage

\section{Introduction}

\IEEEPARstart{S}{ynthetic} aperture radar (SAR) is an active coherent imaging technology that provides high-resolution observations with little dependence on daylight or weather conditions \cite{Moreira2013}. A complex SAR image preserves both the magnitude and phase of the backscattered signal. The magnitude describes scattering strength, while the phase contains information about the propagation path \cite{Moreira2013,Rosen2000}. For two coregistered observations of the same scene, their pointwise conjugate product forms an interferogram. Its argument gives the differential phase, which records the path-length change between the first and second observation epochs. After phase contributions from topography, orbital errors, and atmospheric delays have been modeled and compensated, the remaining differential phase can be related to surface displacement along the radar line of sight \cite{Massonnet1998}. Bitemporal SAR is therefore widely used for deformation monitoring, disaster assessment, and phase-based change analysis. Accurate differential-phase recovery is essential for obtaining reliable measurements in these applications.

In operational SAR systems, complete aperture measurements are often unavailable because acquisition time, onboard data capacity, duty cycle, and operating constraints limit the number and continuity of recorded pulses \cite{Ender2010}. The retained aperture samples may be sparse, nonuniform, or separated by gaps. Periodically gapped raw data can violate the assumptions of conventional frequency-domain focusing and produce severe reconstruction artifacts \cite{Qian2018}. Missing or irregular samples also reshape the effective point spread function, increase sidelobe levels, and introduce spectral aliasing or azimuth ambiguities \cite{Dong2015,Villano2017}. Since SAR image formation coherently integrates complex echoes, these degradations affect both image magnitude and reconstructed phase. Phase errors from either epoch then enter the differential phase and reduce the reliability of subsequent interferometric measurements.

The sampling pattern determines which echo measurements are available for image formation. Conventional SAR acquisition usually adopts uniform slow-time sampling, which provides a regular Doppler grid and supports standard frequency-domain focusing \cite{Cumming2005}. Compressed SAR methods later introduced random or pseudorandom sampling to reduce the number of measurements while retaining sufficient information for sparse scene recovery \cite{Patel2010,Potter2010}. Yang et al. combined Poisson-disk sampling with iterative shrinkage thresholding to support high-resolution and wide-swath imaging \cite{Yang2019}. Villano et al. proposed staggered SAR, in which continuous pulse repetition interval variation distributes blind ranges and reduces concentrated ambiguity energy \cite{Villano2014}. Learning-based methods provide further flexibility. Aggarwal and Jacob demonstrated that sampling locations and reconstruction parameters can be optimized jointly within a model-based network \cite{Aggarwal2020}. In SAR imaging, Zhao et al. learned continuous sampling locations together with a reconstruction network \cite{Zhao2022}, while Wu et al. developed MF-JMoDL-Net to optimize azimuth sampling positions for ambiguity suppression \cite{Wu2024}. These methods improve the point spread function, ambiguity behavior, 
or single-epoch reconstruction quality. Whether such single-epoch 
gains transfer to bitemporal differential phase, however, has not 
been examined.

Beyond the sampling pattern, the reconstruction method determines how faithfully the complex scene is recovered from the retained measurements. Traditional approaches formulate incomplete-data SAR imaging as a regularized inverse problem. Cetin and Karl developed non-quadratic regularization for feature-enhanced complex SAR image formation \cite{Cetin2001}, and Cetin et al. later reviewed sparsity-driven reconstruction, autofocus, and compressed sensing methods \cite{Cetin2014}. Focsa et al. combined compressive sensing with back-projection and demonstrated recovery from fewer measurements while preserving target amplitude, phase, and position \cite{Focsa2021}. Liu et al. constructed a hybrid-domain model and a fast iterative shrinkage algorithm for low-oversampled staggered SAR \cite{Liu2022}. Jiang et al. used motion-parameter estimation, phase compensation, and compressed sensing to reconstruct moving targets from azimuth-missing data \cite{Jiang2022}. Although these methods retain a clear connection to the sensing model, repeated forward and adjoint operations can be computationally demanding. 

Data-driven methods instead learn scene representations or inverse mappings from training data. Yonel et al. unfolded proximal-gradient iterations for passive SAR reconstruction \cite{Yonel2018}, while Pu used an autoencoder formulation for imaging from downsampled echoes and for motion compensation \cite{Pu2021}. Deep unfolding combines trainable priors with explicit data consistency \cite{Monga2021,Wang2025}. Kang et al. unfolded ISTA with an approximate SAR observation operator \cite{Kang2022}, and Wei et al. mapped an ADMM solver into a trainable parametric super-resolution network \cite{Wei2021}. Li et al. developed a matrix-free ADMM network for target-oriented imaging \cite{Li2022}, whereas An et al. incorporated joint low-rank and sparse priors into an unfolded recovery network \cite{An2022}. This balance between physical structure, learned regularization, and predictable inference makes deep unfolding a suitable foundation for complex SAR reconstruction with incomplete measurements.

Despite these advances, improvements in the point spread function, ambiguity suppression, or single-epoch image quality do not necessarily produce accurate bitemporal differential phase. Most reconstruction losses and evaluation metrics are applied to each epoch separately and mainly emphasize magnitude fidelity. They therefore provide no direct constraint on the relative phase between corresponding complex pixels. The differential phase is obtained from the argument of the conjugate product of the two reconstructed SAR images, so phase errors in either epoch propagate into the final interferogram \cite{Bamler1998}. Even when both magnitude images are well focused and achieve high PSNR or SSIM, a small spatially varying phase bias may reduce local coherence or distort deformation estimates \cite{Zebker1992}. Studies of coherent inverse problems further show that magnitude-based regularization must be designed carefully when the phase needs to be preserved \cite{Cetin2006,Watson2025}. Related research has addressed interferometric phase restoration after image formation. Deledalle et al. proposed NL-InSAR for nonlocal estimation of reflectivity, coherence, and interferometric phase \cite{Deledalle2011}, while Sica et al. developed InSAR-BM3D for structure-preserving phase filtering \cite{Sica2018}. Sun et al. introduced DeepInSAR to jointly estimate restored phase and coherence \cite{Sun2020}. Ding et al. subsequently incorporated coherence guidance into complex convolutional sparse coding \cite{Ding2022} and improved its robustness to outliers \cite{Ding2024}. These methods operate on already formed interferograms or SLC pairs. They do not determine which raw echo measurements should be acquired or how the two complex images should be reconstructed for the differential-phase task.

The sampling pattern determines which parts of the complex echo are retained, while the reconstruction network determines how magnitude and phase are inferred from those measurements. Both should therefore be assessed under the differential-phase objective rather than under single-epoch criteria alone.

Motivated by this observation, this paper proposes the 
Differential-Phase-Oriented Joint Masked Reconstruction Network (DP-JMRNet), a physics-guided framework for complex bitemporal SAR reconstruction under a differential-phase objective. The acquisition side adopts a standard aperture pattern that satisfies a prescribed sampling budget, 
a maximum-gap constraint, and a shared support across the two epochs; the role and limits of pattern design are examined empirically in Section~\ref{sec:results}. For reconstruction, a complex-valued unfolding network alternates matrix-free SAR data-consistency updates with learned complex priors. An exchange-equivariant selective interaction module 
uses reliability gating to adaptively exploit complementary 
information from the two epochs. The main contributions of this work 
are summarized as follows.

\begin{enumerate}

    \item \textit{Differential-phase-oriented bitemporal reconstruction:}
This paper introduces DP-JMRNet, a physics-guided framework in which 
a shared aperture mask and a complex-valued unfolding network operate 
jointly on the two epochs under a differential-phase objective. 
Existing sparse SAR methods are supervised by single-epoch magnitude 
or complex criteria; here the training objective includes the 
differential phase itself, so that the reconstruction serves the final 
interferometric task. To the best of our knowledge, this is the first 
sparse SAR reconstruction network supervised directly by a 
differential-phase criterion.

    \item \textit{Reliable bitemporal unfolding with selective interaction:}
The reconstructor is a complex-valued unfolding network that alternates 
matrix-free SAR data-consistency updates with learned complex priors, 
augmented by two components that govern cross-epoch information flow. 
An exchange-equivariant interaction module ensures that swapping the 
two epochs swaps the outputs exactly, removing any dependence on epoch 
ordering, verified to a swap error of $10^{-9}$. A coherence-aware 
reliability gate promotes information exchange in coherent regions 
while suppressing unreliable interaction in low-coherence or 
conflicting areas. Ablations show that these components improve both 
differential-phase accuracy and single-epoch amplitude reconstruction, 
indicating that they extract complementary scene information rather 
than redistributing phase error.

\item \textit{Characterization of acquisition design limits:}
This paper establishes the feasibility constraints required for 
phase-preserving bitemporal acquisition and delimits what remains 
achievable within them. A bounded maximum aperture gap excludes 
periodic decimation, whose grating lobes produce strong ambiguities. Within the resulting feasible set, a cumulative hill-climbing search 
with progressively larger deviation budgets yields no measurable 
improvement on held-out scenes. This establishes that for a fixed reconstructor, 
no aperture pattern substantially improves the differential phase.

    \item \textit{Comprehensive experimental validation:}
    Experiments are conducted at three sampling rates on a simulated bitemporal SAR dataset and on external real Sentinel-1 SLC data \cite{Torres2012}. Comparisons with representative iterative and learning-based methods, together with ablation, robustness, complexity, and qualitative analyses, demonstrate that DP-JMRNet improves differential-phase recovery while maintaining competitive complex-image reconstruction quality.
\end{enumerate}

\section{Problem Formulation}
\label{sec:problem_formulation}

\subsection{Bitemporal SAR Observation Model}
\label{subsec:observation_model}

Consider two coregistered complex SAR images acquired over the same scene at
two observation epochs. Let $x_{t}\in\mathbb{C}^{H\times W}$ denote the complex
reflectivity image at epoch $t\in\{1,2\}$, vectorized as an element of
$\mathbb{C}^{N}$ with $N=HW$. After system calibration, motion compensation,
and reference-range demodulation, the full complex phase history can be
expressed as
\begin{equation}
    s_{t}=Ax_{t},
    \qquad t\in\{1,2\},
    \label{eq:full_phase_history}
\end{equation}
where $A:\mathbb{C}^{N}\rightarrow\mathbb{C}^{N_{a}N_{f}}$ is the SAR forward
operator, $N_{a}$ is the number of candidate aperture positions, and $N_{f}$ is
the number of frequency samples at each position. The operator follows the
reference-demodulated monostatic far-field model, with deterministic waveform
and geometric factors absorbed into $A$ \cite{Cumming2005}; its specific
parameters are given in Section~\ref{sec:experimental_setup}. Because the two epochs
share the same nominal acquisition geometry, a single operator $A$ is used for
both.

Let $M\in\{0,1\}^{N_{a}}$ denote a binary aperture mask, applied in the
physically meaningful phase-history domain and broadcast over the frequency
dimension. An entry $M[n]=1$ indicates that all $N_{f}$ frequency samples at
the $n$th aperture position are retained. Otherwise, that aperture position is
omitted. If $K$ positions are selected, then
\begin{equation}
    \lVert M\rVert_{0}=K,
    \qquad
    \rho=\frac{K}{N_{a}},
    \label{eq:sampling_budget}
\end{equation}
where $\rho$ is the aperture sampling rate. Defining the masked forward
operator $A_{M}=MA$, the undersampled observations are given by
\begin{equation}
    y_{t}
    =
    M\left(Ax_{t}+\nu_{t}\right)
    =
    A_{M}x_{t}+n_{t},
    \qquad t\in\{1,2\},
    \label{eq:undersampled_observation}
\end{equation}
where $\nu_{t}$ is noise on the full phase-history grid and $n_{t}=M\nu_{t}$ is
the retained measurement noise. We assume independent circular complex Gaussian
noise,
\begin{equation}
    n_{t}
    \sim
    \mathcal{CN}
    \left(
        0,
        \sigma_{t}^{2}I
    \right).
    \label{eq:noise_model}
\end{equation}

The two epochs use one shared aperture mask,
\begin{equation}
    M_{1}
    =
    M_{2}
    =
    M.
    \label{eq:shared_mask}
\end{equation}
This shared support is not merely a modeling convenience. As shown in
Section~\ref{sec:results}, breaking the pairing between epochs degrades the
differential phase substantially even when the sampling budget is unchanged.
Nevertheless, the measured values and noise realizations remain epoch dependent
because $x_{1}$ and $x_{2}$ describe different complex scattering states.

\subsection{Reconstruction and Differential-Phase Task}
\label{subsec:differential_phase_task}

Given the two undersampled observations, a bitemporal reconstruction model with
parameters $\Theta$ produces two complex-valued estimates:
\begin{equation}
    \left(
        \hat{x}_{1},
        \hat{x}_{2}
    \right)
    =
    R_{\Theta}
    \left(
        y_{1},
        y_{2};
        A,
        M
    \right).
    \label{eq:bitemporal_reconstruction}
\end{equation}
Unlike magnitude-only reconstruction, $R_{\Theta}$ must preserve both the
scattering magnitude and the phase of each image.

For a pixel $p$, the reference differential phase is defined using the
conjugate product
\begin{equation}
    \Delta\phi[p]
    =
    \operatorname{wrap}
    \left(
        \angle
        \left(
            x_{2}[p]x_{1}^{*}[p]
        \right)
    \right)
    =
    \operatorname{wrap}
    \left(
        \phi_{2}[p]-\phi_{1}[p]
    \right),
    \label{eq:reference_differential_phase}
\end{equation}
where $\phi_{t}[p]=\angle x_{t}[p]$ and
$\operatorname{wrap}(\alpha)=\operatorname{atan2}(\sin\alpha,\cos\alpha)
\in(-\pi,\pi]$. The reconstructed differential phase is obtained in the same
manner:
\begin{equation}
    \widehat{\Delta\phi}[p]
    =
    \operatorname{wrap}
    \left(
        \angle
        \left(
            \hat{x}_{2}[p]
            \hat{x}_{1}^{*}[p]
        \right)
    \right).
    \label{eq:estimated_differential_phase}
\end{equation}
This convention gives the phase change from the first epoch to the second
epoch. Reversing the complex product would reverse the sign and is therefore
not used. After nuisance-phase compensation and phase unwrapping, the
differential phase maps linearly to the line-of-sight range change
\cite{Bamler1998}.

Interferometric coherence indicates whether the phase difference is locally
reliable. Let $\Omega_{p}$ be a neighborhood centered at pixel $p$, and let
$h_{p,q}\geq 0$ denote normalized spatial weights satisfying
$\sum_{q\in\Omega_{p}}h_{p,q}=1$. The local coherence magnitude is defined as
\begin{equation}
    \gamma[p]
    =
    \frac{
        \left|
        \sum_{q\in\Omega_{p}}
        h_{p,q}
        x_{2}[q]x_{1}^{*}[q]
        \right|
    }{
        \sqrt{
            \left(
                \sum_{q\in\Omega_{p}}
                h_{p,q}|x_{1}[q]|^{2}
            \right)
            \left(
                \sum_{q\in\Omega_{p}}
                h_{p,q}|x_{2}[q]|^{2}
            \right)
        }
        +\varepsilon
    },
    \label{eq:local_coherence}
\end{equation}
where $\varepsilon>0$ prevents numerical instability and
$\gamma[p]\in[0,1]$ \cite{Deledalle2011}. The estimated coherence
$\hat{\gamma}[p]$ is obtained by replacing $x_{1}$ and $x_{2}$ in
\eqref{eq:local_coherence} with their reconstructions.

To show how reconstruction errors enter the interferometric result, define the
single-epoch phase errors as
\begin{equation}
    e_{t}[p]
    =
    \operatorname{wrap}
    \left(
        \angle
        \left(
            \hat{x}_{t}[p]x_{t}^{*}[p]
        \right)
    \right),
    \qquad t\in\{1,2\}.
    \label{eq:single_epoch_phase_error}
\end{equation}
On pixels with nonzero magnitude, the differential-phase error satisfies
\begin{align}
    e_{\Delta\phi}[p]
    &=
    \operatorname{wrap}
    \left(
        \widehat{\Delta\phi}[p]
        -
        \Delta\phi[p]
    \right)
    \nonumber\\
    &=
    \operatorname{wrap}
    \left(
        e_{2}[p]-e_{1}[p]
    \right).
    \label{eq:differential_phase_error}
\end{align}
Thus, a phase error in either reconstructed image directly affects the final
interferogram, and only the difference between the two errors is observable.
Error components that are common to both epochs cancel, whereas unequal or
spatially varying components remain. This is the same mechanism by which
orbital, topographic, and atmospheric phase terms cancel in conventional
repeat-pass interferometry. Because both epochs are observed through the same
aperture support in \eqref{eq:shared_mask}, sampling-induced errors are
largely common mode and are therefore attenuated in the differential phase,
whereas errors driven by epoch-specific noise and scattering are not. This also
explains why low magnitude error or high single-epoch PSNR does not by itself
guarantee an accurate differential phase.

Because phase is poorly defined in low-amplitude or decorrelated regions, we
introduce a reliability map with entries $w_{p}\in[0,1]$. It can incorporate
coherence, amplitude support, and other validity information. The primary
differential-phase error is measured using a reliability-weighted circular
RMSE:
\begin{equation}
    \mathcal{L}_{\Delta\phi}
    =
    \left[
    \frac{
        \sum_{p=1}^{N}
        w_{p}\,
        e_{\Delta\phi}^{2}[p]
    }{
        \sum_{p=1}^{N}w_{p}+\varepsilon
    }
    +\varepsilon
    \right]^{1/2}.
    \label{eq:differential_phase_loss}
\end{equation}
The wrapped error in \eqref{eq:differential_phase_error} avoids the artificial
discontinuity between phase values close to $-\pi$ and $\pi$.

\subsection{Feasible Sampling and Reconstruction Objective}
\label{subsec:joint_problem}

The sampling mask must satisfy both the measurement budget and an
aperture-continuity constraint. Let $a_{1}<a_{2}<\cdots<a_{K}$ denote the
selected aperture indices. The maximum number of consecutive missing positions
is
\begin{equation}
    \operatorname{gap}(M)
    =
    \max
    \left\{
        a_{1}-1,\,
        \max_{1\leq k<K}
        \left(a_{k+1}-a_{k}-1\right),\,
        N_{a}-a_{K}
    \right\}.
    \label{eq:maximum_gap}
\end{equation}
For a prescribed maximum gap $g$, the feasible-mask set is
\begin{equation}
    \mathcal{M}_{K,g}
    =
    \left\{
        M\in\{0,1\}^{N_{a}}
        \,\middle|\,
        \lVert M\rVert_{0}=K,\,
        \operatorname{gap}(M)\leq g
    \right\}.
    \label{eq:feasible_mask_set}
\end{equation}
This constraint serves two purposes. It prevents a mask from satisfying the
global sampling rate while leaving an excessively long unsampled aperture
interval, and it excludes strictly periodic decimation, whose grating lobes
produce strong azimuth ambiguities.

In addition to the differential-phase loss, complex and magnitude
reconstruction losses are included to preserve the physical content of both
epochs. Let
\begin{equation}
    D
    =
    \sum_{t=1}^{2}
    \lVert x_{t}\rVert_{2}^{2}
    +\varepsilon.
    \label{eq:loss_normalization}
\end{equation}
The normalized complex and amplitude losses are defined as
\begin{align}
    \mathcal{L}_{\mathrm{complex}}
    &=
    \frac{
        \sum_{t=1}^{2}
        \lVert
            \hat{x}_{t}
            -
            x_{t}
        \rVert_{2}^{2}
    }{D},
    \label{eq:complex_reconstruction_loss}\\
    \mathcal{L}_{\mathrm{amp}}
    &=
    \frac{
        \sum_{t=1}^{2}
        \left\lVert
            |\hat{x}_{t}|
            -
            |x_{t}|
        \right\rVert_{2}^{2}
    }{D}.
    \label{eq:amplitude_reconstruction_loss}
\end{align}
A reliability-weighted single-epoch phase loss is also used:
\begin{equation}
    \mathcal{L}_{\mathrm{raw}\phi}
    =
    \left[
        \frac{1}{2}
        \sum_{t=1}^{2}
        \frac{
            \sum_{p=1}^{N}
            w_{p}\,e_{t}^{2}[p]
        }{
            \sum_{p=1}^{N}w_{p}+\varepsilon
        }
        +\varepsilon
    \right]^{1/2}.
    \label{eq:single_epoch_phase_loss}
\end{equation}
Both phase losses are circular and reliability weighted, so low-amplitude or
low-coherence pixels do not receive the same influence as reliable
interferometric pixels. The complete reconstruction objective is
\begin{equation}
\begin{split}
    \mathcal{L}_{\mathrm{rec}}
    ={}&
    \lambda_{c}\mathcal{L}_{\mathrm{complex}}
    +
    \lambda_{a}\mathcal{L}_{\mathrm{amp}} \\
    &+
    \lambda_{r}\mathcal{L}_{\mathrm{raw}\phi}
    +
    \lambda_{\Delta}\mathcal{L}_{\Delta\phi},
\end{split}
    \label{eq:reconstruction_objective}
\end{equation}
where the nonnegative coefficients balance complex reconstruction, magnitude
fidelity, single-epoch phase preservation, and differential-phase accuracy. In
particular, $\lambda_{\Delta}>0$ makes the final interferometric quantity part
of the optimization objective. The other terms prevent a phase-only solution
from sacrificing the physical quality of the reconstructed SAR images.

Given a mask $M\in\mathcal{M}_{K,g}$ and the corresponding observations in
\eqref{eq:undersampled_observation}, the reconstruction parameters are obtained
by
\begin{equation}
    \Theta^{\star}
    =
    \underset{\Theta}{\operatorname*{arg\,min}}
    \;
    \mathbb{E}
    \left[
        \mathcal{L}_{\mathrm{rec}}
        \left(
            R_{\Theta}(y_{1},y_{2};A,M),\,
            X
        \right)
    \right],
    \label{eq:reconstruction_optimization}
\end{equation}
where $X=(x_{1},x_{2})$ collects the reference images and the expectation is
taken over bitemporal scenes and measurement noise.

The mask itself is not optimized. It is drawn from a standard aperture pattern
family satisfying \eqref{eq:feasible_mask_set}, and the same mask is shared by
both epochs during training and evaluation. Whether pattern design within
$\mathcal{M}_{K,g}$ can further improve the differential phase is examined
empirically in Section~\ref{sec:results}.

\section{Proposed Method}
\label{sec:proposed_method}

\subsection{Overall Framework of DP-JMRNet}
\label{subsec:overall_framework}

\begin{figure*}[t]
    \centering
    \includegraphics[width=\textwidth]{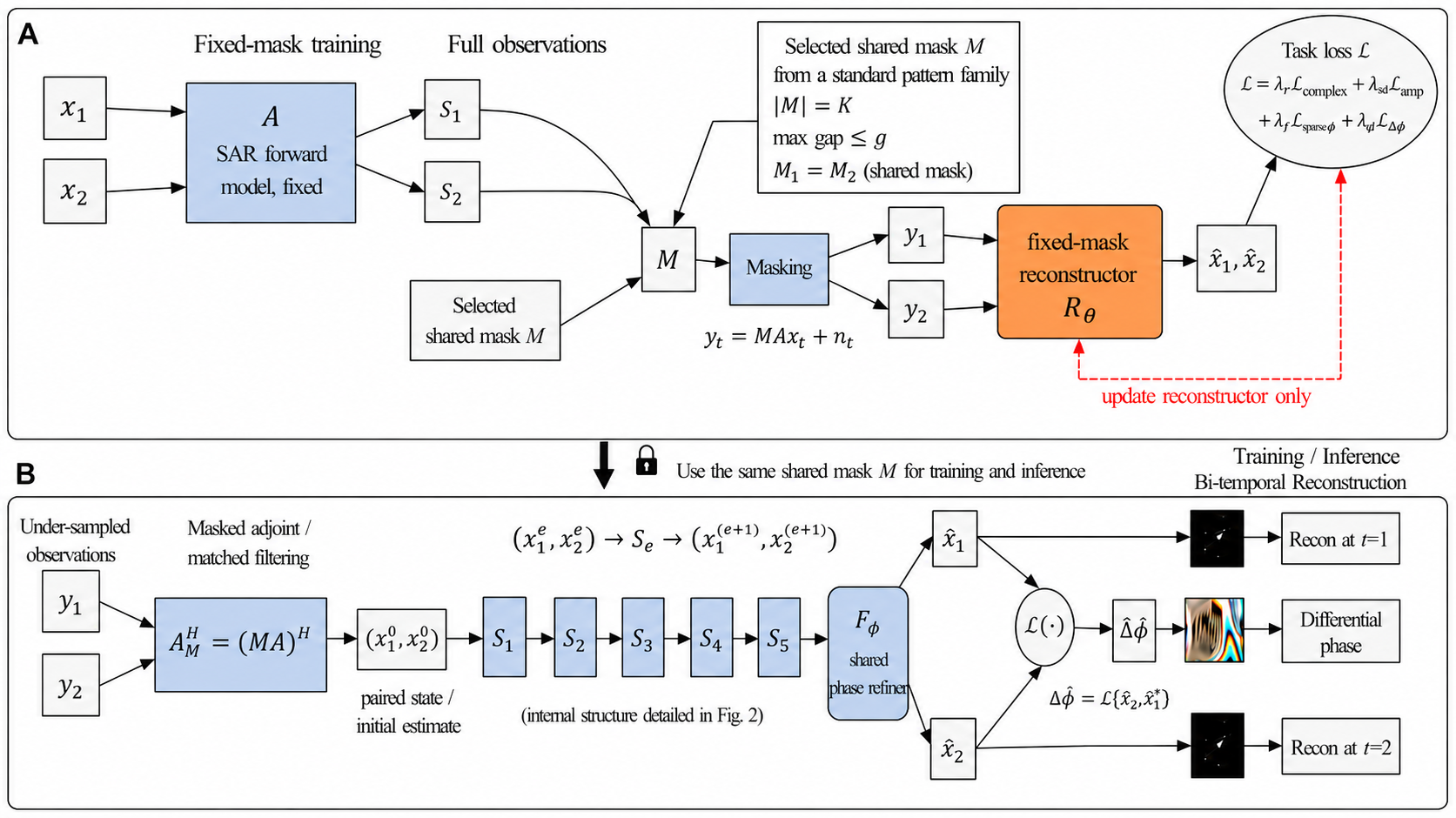}
    \caption{Overall workflow of DP-JMRNet. (A) Fixed Mask Training Process.
    (B) Bi-Temporal Reconstruction.}
    \label{fig:proposed_overall}
\end{figure*}

As illustrated in Fig.~\ref{fig:proposed_overall}, the overall framework of
DP-JMRNet has two closely matched parts. During training, a physically feasible
aperture mask is kept fixed while the reconstruction parameters are optimized
from paired complex SAR scenes. During inference, the same mask and its matched
reconstructor are used without mask adaptation. Fixing the mask ensures that
the training and inference operators are identical, and that both epochs are
observed through the same aperture support as required by
\eqref{eq:shared_mask}.

\subsubsection{Fixed shared-mask setting}
The observation model of \eqref{eq:undersampled_observation} is used throughout,
with the shared mask fixed to a feasible pattern $M\in\mathcal{M}_{K,g}$ drawn
from a standard aperture pattern family. The mask is stored as a frozen binary
array and is excluded from the trainable parameter set, so no gradient or
optimizer update is applied to it. M is fixed and non-trainable. It does not receive gradient updates, but it is provided as a fixed conditioning variable to the reconstruction modules. For each sampling budget,
the corresponding mask and reconstructor form a matched pair.

\subsubsection{Paired reconstruction path}
In \eqref{eq:bitemporal_reconstruction}, $\Theta$ collects all trainable
parameters of the proposed reconstructor, including the unfolding priors,
selective interaction modules, step sizes, the mask conditioner, and the final
phase refiner. The semicolon in that expression emphasizes that $A$ and $M$ are
fixed physical conditions rather than optimized variables.

The reconstruction begins with an independent matched-filtering initialization
for each temporal track. Since
\begin{equation}
    A_{M}^{H}=(MA)^{H},
    \label{eq:masked_adjoint}
\end{equation}
the initial paired state is
\begin{equation}
    x_{t}^{(0)}=A_{M}^{H}y_{t},
    \qquad t\in\{1,2\}.
    \label{eq:adjoint_initialization}
\end{equation}
Starting from $(x_{1}^{(0)},x_{2}^{(0)})$, the network applies 5 unfolding
stages. To avoid ambiguity between the stage label and the state index, we
collect the two channels into $\mathbf{x}^{(\ell)}\triangleq
\bigl(x_{1}^{(\ell)},x_{2}^{(\ell)}\bigr)$ and
$\mathbf{y}\triangleq(y_{1},y_{2})$, so that the cascade reads
\begin{equation}
    \mathbf{x}^{(\ell)}
    =\mathcal{S}_{\ell}\bigl(\mathbf{x}^{(\ell-1)};
     \mathbf{y},A,M\bigr),
    \qquad \ell=1,\ldots,L,
    \label{eq:stage_cascade}
\end{equation}
where the superscript $(\ell)$ denotes the stage index and the subscripts
$1,2$ denote the two acquisitions. Each stage performs independent pre-prior
data consistency, a complex prior shared between the temporal tracks,
coherence-aware selective interaction, and independent post-interaction data
consistency. The internal construction of $\mathcal{S}_{\ell}$ is detailed in
Section~\ref{subsec:single_stage} and Fig.~\ref{fig:single_stage}.

After the fifth stage, a single phase-refinement network $F_{\phi}$ is reused
for both tracks. Let
\begin{equation}
    q_{t}=\tanh\!\left[
        F_{\phi}\!\left(
            [\operatorname{Re}(x_{t}^{(L)}),
             \operatorname{Im}(x_{t}^{(L)})]
        \right)
    \right]
    \label{eq:phase_refinement_map}
\end{equation}
be its bounded phase-correction map, and let $\eta_{t}(M)\in[-1,1]$ be the
track-wise coefficient produced by the shared mask conditioner, which takes the
fixed mask as a conditioning input without modifying it. The final output is
\begin{equation}
    \delta_{t}=0.08\,\eta_{t}(M)q_{t},
    \qquad
    \hat{x}_{t}=x_{t}^{(L)}\exp(j\delta_{t}).
    \label{eq:phase_only_refinement}
\end{equation}
The scaling constant restricts the phase correction to
$|\delta_{t}|\leq 0.08$~rad, so that the refinement can only make small
adjustments and cannot override the phase produced by the unfolding stages.
Thus, the last module changes only phase and leaves the reconstructed amplitude
unchanged. Because the same $F_{\phi}$ and mask conditioner are applied to both
tracks, this refinement does not introduce epoch-specific parameters. The
reconstructed wrapped differential phase then follows
\eqref{eq:estimated_differential_phase}.

\subsubsection{Training objective}
Only $\Theta$ is optimized during the training process shown in
Fig.~\ref{fig:proposed_overall}(A). The reconstruction objective is
\eqref{eq:reconstruction_objective}, to which the two gate regularizers
introduced with the single-stage model are added:
\begin{equation}
    \mathcal{L}_{\mathrm{train}}
    =\mathcal{L}_{\mathrm{rec}}
     +\lambda_{s}\mathcal{R}_{\mathrm{smooth}}
     +\lambda_{\mathrm{ac}}\mathcal{R}_{\mathrm{anti\text{-}collapse}},
    \label{eq:complete_training_objective}
\end{equation}
where $\mathcal{R}_{\mathrm{smooth}}$ and
$\mathcal{R}_{\mathrm{anti\text{-}collapse}}$ are defined in
Section~\ref{subsec:single_stage}. Fixed-mask learning therefore solves
\begin{equation}
    \Theta^{\star}
    =\arg\min_{\Theta}\;
    \mathbb{E}\!\left[\mathcal{L}_{\mathrm{train}}\right],
    \label{eq:fixed_mask_optimization}
\end{equation}
and no gradient or optimizer update is applied to $M$. This gives the same
observation operator, the same aperture support, and the same reconstruction
path during training and inference.

\subsection{Exchange-Equivariant Selective Reconstruction}
\label{subsec:single_stage}

\begin{figure*}[t]
    \centering
    \includegraphics[width=\textwidth]{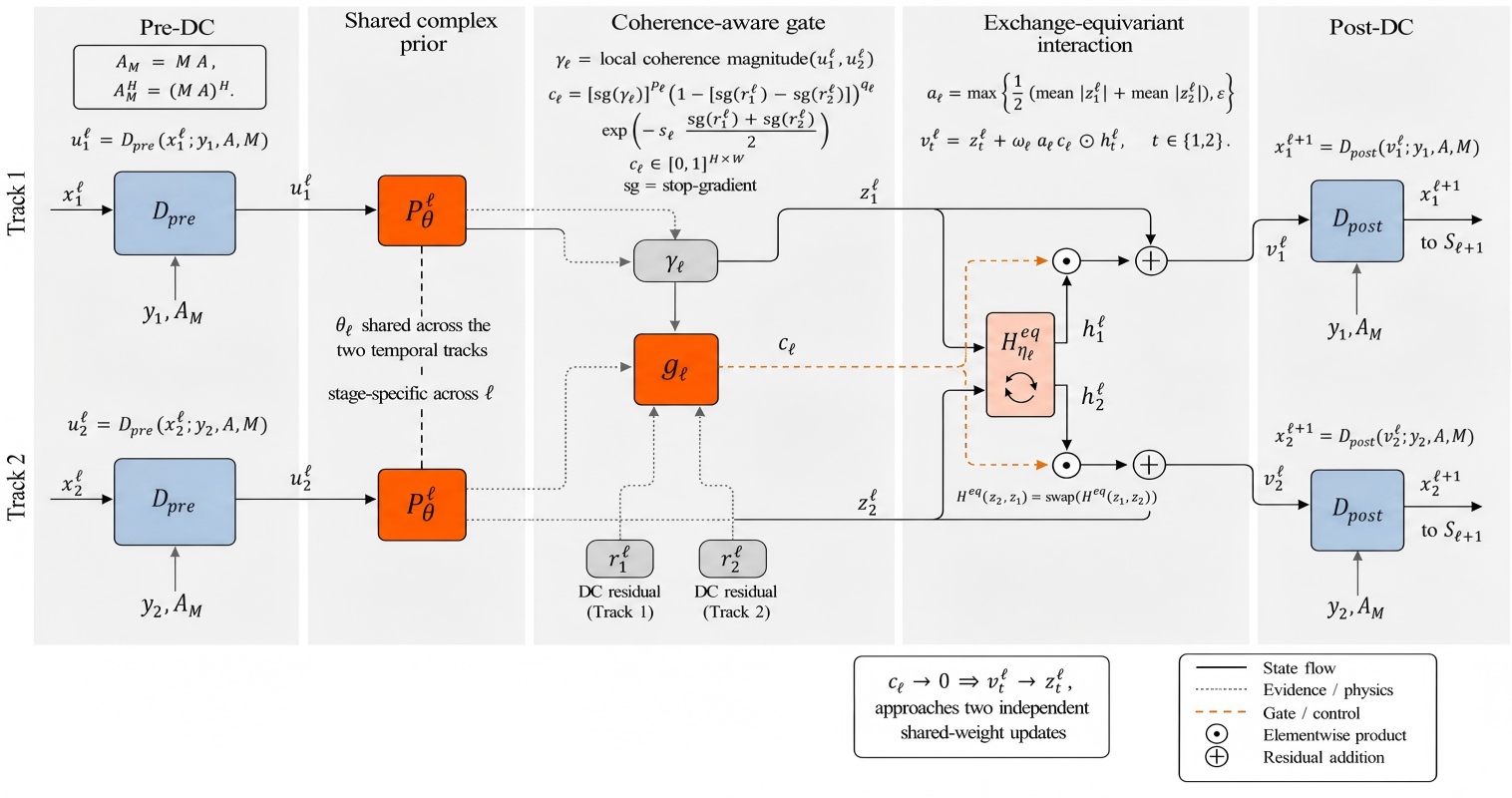}
    \caption{Internal structure of one exchange-equivariant selective
    unfolding stage in DP-JMRNet.}
    \label{fig:single_stage}
\end{figure*}

Figure~\ref{fig:single_stage} details one unfolding stage of DP-JMRNet.
For $\ell=1,\ldots,L$, stage $\mathcal{S}_{\ell}$ maps the paired state
$(x_{1}^{(\ell-1)},x_{2}^{(\ell-1)})$ to
$(x_{1}^{(\ell)},x_{2}^{(\ell)})$. Each stage contains pre-prior data
consistency, a temporally shared complex prior, coherence-aware selective
interaction, and post-interaction data consistency, shown from left to right
as the five blocks of Fig.~\ref{fig:single_stage}. The learned operators are
stage specific, whereas their parameters are shared between the two temporal
tracks. All trainable parameters introduced below are collected in the global
parameter set $\Theta$.

\subsubsection{Data consistency and shared complex prior}

The Pre-DC, Shared complex prior, and Post-DC blocks of
Fig.~\ref{fig:single_stage} form the backbone of a stage, with the two tracks
drawn as the upper and lower rows. Let $A_{M}=MA$. We use the compact
data-consistency operator
\begin{equation}
    \mathcal{D}_{\alpha}(w;y_{t},A,M)
    =w-\alpha A_{M}^{H}(A_{M}w-y_{t}).
    \label{eq:dc_operator}
\end{equation}
The pre-prior update of track $t\in\{1,2\}$ is
\begin{equation}
    u_{t}^{(\ell)}
    =\mathcal{D}_{\alpha_{\mathrm{pre}}^{(\ell)}}
     \bigl(x_{t}^{(\ell-1)};y_{t},A,M\bigr),
    \label{eq:pre_dc}
\end{equation}
which is followed by the shared complex prior
\begin{equation}
    z_{t}^{(\ell)}
    =P_{\ell}\bigl(u_{t}^{(\ell)};M\bigr),
    \label{eq:shared_prior}
\end{equation}
and, after the interaction step, by the post-interaction update
\begin{equation}
    x_{t}^{(\ell)}
    =\mathcal{D}_{\alpha_{\mathrm{post}}^{(\ell)}}
     \bigl(v_{t}^{(\ell)};y_{t},A,M\bigr),
    \label{eq:post_dc}
\end{equation}
with $v_{t}^{(\ell)}$ given in \eqref{eq:gated_update}. Here $P_{\ell}$ is a
stage-specific complex residual prior applied with the same parameters to both
epochs. The shared prior imposes a common reconstruction rule without forcing
the two temporal states to be identical. The pre and post-interaction updates
use the observation of the corresponding track, so information exchange does
not replace measurement consistency.

\subsubsection{Coherence-aware selective gate}

The Coherence-aware gate block of Fig.~\ref{fig:single_stage} produces a
single spatial gate that is fed to both tracks. The gate is constructed from
two magnitude-valued sources of evidence. The first is the local coherence
magnitude $\gamma^{(\ell)}$, computed as in \eqref{eq:local_coherence} with
$x_{1}$ and $x_{2}$ replaced by $u_{1}^{(\ell)}$ and $u_{2}^{(\ell)}$. The
second is the normalized image-domain data-consistency gradient
$d_{t}^{(\ell)}=A_{M}^{H}(A_{M}x_{t}^{(\ell-1)}-y_{t})$, from which the
residual evidence is
\begin{equation}
\begin{aligned}
    \chi_{t}^{(\ell)}
    &=\frac{\bigl|d_{t}^{(\ell)}\bigr|}
      {\sqrt{\operatorname{mean}
      \bigl|d_{t}^{(\ell)}\bigr|^{2}+\varepsilon}},\\
    r_{t}^{(\ell)}
    &=\frac{\chi_{t}^{(\ell)}}{1+\chi_{t}^{(\ell)}}.
\end{aligned}
\label{eq:residual_evidence}
\end{equation}
The resulting gate is
\begin{equation}
\begin{split}
    c^{(\ell)}
    ={}&\bigl[\operatorname{sg}(\gamma^{(\ell)})\bigr]^{p_{\ell}}
      \bigl[1-\bigl|\operatorname{sg}(r_{1}^{(\ell)})
      -\operatorname{sg}(r_{2}^{(\ell)})\bigr|\bigr]^{q_{\ell}}\\
    &\times
      \exp\!\left[-\frac{s_{\ell}}{2}
      \operatorname{sg}\!\bigl(r_{1}^{(\ell)}+r_{2}^{(\ell)}\bigr)\right],
\end{split}
\label{eq:gate_definition}
\end{equation}
where $\operatorname{sg}(\cdot)$ denotes stop-gradient and
$c^{(\ell)}\in[0,1]^{H\times W}$. The coherence term favors reliable common
structure, the disagreement term suppresses conflicting track-wise evidence,
and the exponential term attenuates interaction where the mean residual is
large. Because all evidence enters symmetrically, exchanging the two epochs
does not change the gate. The first-stage gate is analytically anchored,
whereas the remaining stages learn positive values of
$p_{\ell}$, $q_{\ell}$, and $s_{\ell}$.

\subsubsection{Exchange-equivariant interaction}

The Exchange-equivariant interaction block of Fig.~\ref{fig:single_stage}
contains a single shared module with two inputs and two outputs, followed by
the gated residual update on each track. The shared interaction operator
produces paired complex corrections
\begin{equation}
    \bigl(h_{1}^{(\ell)},h_{2}^{(\ell)}\bigr)
    =H_{\ell}^{\mathrm{eq}}
     \bigl(z_{1}^{(\ell)},z_{2}^{(\ell)}\bigr),
    \label{eq:interaction_output}
\end{equation}
and is evaluated in opposite endpoint orders, so that
\begin{equation}
    H_{\ell}^{\mathrm{eq}}(z_{2},z_{1})
    =\operatorname{swap}
     \bigl(H_{\ell}^{\mathrm{eq}}(z_{1},z_{2})\bigr),
    \label{eq:exchange_equivariance}
\end{equation}
which preserves temporal exchange equivariance. Exchange equivariance is enforced by construction rather than encouraged by an
auxiliary loss. Let $a_{i}=\bigl|z_{i}^{(\ell)}\bigr|$. A permutation-invariant
pair descriptor is first formed from symmetric statistics of the two magnitudes,
\begin{equation}
    s^{(\ell)}
    =S_{\ell}\!\left(
     \frac{a_{1}+a_{2}}{2},\;
     \bigl|a_{1}-a_{2}\bigr|,\;
     \sqrt{a_{1}a_{2}},\;
     \max(a_{1},a_{2})\right),
    \label{eq:pair_descriptor}
\end{equation}
each argument of which is invariant to exchanging the two epochs. A shared
endpoint encoder $E_{\ell}$ and a shared decoder $D_{\ell}$ are then applied in
the two ordered arrangements,
\begin{equation}
\begin{aligned}
    h_{1}^{(\ell)}
    &=D_{\ell}\!\left(s^{(\ell)},
      E_{\ell}(z_{1}^{(\ell)}),
      E_{\ell}(z_{2}^{(\ell)})\right),\\
    h_{2}^{(\ell)}
    &=D_{\ell}\!\left(s^{(\ell)},
      E_{\ell}(z_{2}^{(\ell)}),
      E_{\ell}(z_{1}^{(\ell)})\right),
\end{aligned}
    \label{eq:ordered_decoding}
\end{equation}
so that the decoder retains endpoint order while all shared components are
order independent. Exchanging the two inputs therefore exchanges the two
outputs exactly, up to floating-point precision.

The corrections are rescaled
by the scene-level factor
\begin{equation}
    a_{\ell}
    =\max\!\left\{
     \frac{\operatorname{mean}\bigl|z_{1}^{(\ell)}\bigr|
     +\operatorname{mean}\bigl|z_{2}^{(\ell)}\bigr|}{2},\,
     \varepsilon\right\},
    \label{eq:scene_scale}
\end{equation}
and the gated residual update of track $t\in\{1,2\}$ is
\begin{equation}
    v_{t}^{(\ell)}
    =z_{t}^{(\ell)}
     +\omega_{\ell}a_{\ell}c^{(\ell)}\odot h_{t}^{(\ell)}.
    \label{eq:gated_update}
\end{equation}
The scene-level scale $a_{\ell}$ makes the correction less sensitive to the
absolute signal magnitude, and $\omega_{\ell}$ controls its stage-wise
strength. The spatial gate acts only on the interaction term; therefore,
$c^{(\ell)}\rightarrow0$ implies
$v_{t}^{(\ell)}\rightarrow z_{t}^{(\ell)}$, continuously recovering two
independent shared-weight updates.

\subsubsection{Gate regularization and stage symmetry}

The gate produced by the Coherence-aware gate block of
Fig.~\ref{fig:single_stage} is additionally constrained during training. As
introduced in the fixed-mask training objective, the gate is regularized to
remain spatially smooth and to avoid a trivial all-zero solution in regions
supported by favorable evidence:
\begin{equation}
    \mathcal{R}_{\mathrm{smooth}}
    =\frac{1}{L}\sum_{\ell=1}^{L}
     \operatorname{mean}\bigl\|\nabla c^{(\ell)}\bigr\|_{2}^{2},
    \label{eq:smoothness_regularizer}
\end{equation}
\begin{equation}
    \mathcal{R}_{\text{anti-collapse}}
    =\frac{1}{L}\sum_{\ell=1}^{L}
     \bigl[\tau-\bar c_{\mathrm{elig}}^{(\ell)}\bigr]_{+}^{2},
    \label{eq:anti_collapse_regularizer}
\end{equation}
where $\bar c_{\mathrm{elig}}^{(\ell)}$ is the evidence-weighted mean gate
activation over high-coherence, low-residual locations. The corresponding
thresholds and loss weights are reported in the experimental setup.

The data-consistency maps and learned operators are shared across the two
tracks, the gate is symmetric, and $H_{\ell}^{\mathrm{eq}}$ is exchange
equivariant, as annotated in Fig.~\ref{fig:single_stage}. Consequently,
swapping the two inputs and their observations only swaps the two stage
outputs. This property is preserved by composing all $L$ stages of DP-JMRNet.


\section{Experimental Setup}
\label{sec:experimental_setup}

\subsection{Datasets}
\label{subsec:datasets}

\subsubsection{Bitemporal simulated SAR dataset}

We construct a bitemporal simulated SAR dataset containing 4000 pairs of
$256\times256$ complex-valued images as our main dataset. Instead of enlarging
low-resolution images, every parent scene is simulated directly on a
$1024\times1024$ grid. At most sixteen non-overlapping patches are extracted
from each scene at the original resolution. Each parent combines a distributed
background, 5--9 extended scattering regions, 3--7 linear structures, and 5--10
clusters of point scatterers. Spatially varying phase is assigned to the first
epoch so that the complex reference is not reduced to an amplitude image with a
constant phase.

The second epoch is generated from the same underlying scene with two to five
smooth local deformation regions, local amplitude modulation, scatterer
appearance or disappearance, and one to three regions of reduced coherence. The
simulated changes are divided into easy, medium, and hard cases with
probabilities 25\%, 50\%, and 25\%, respectively. Their ranges are listed in
Table~\ref{tab:simulated_dataset_difficulty}.

\begin{table}[!ht]
\centering
\caption{Difficulty Ranges of the Bitemporal Simulated SAR Dataset}
\label{tab:simulated_dataset_difficulty}
\footnotesize
\setlength{\tabcolsep}{4.5pt}
\begin{tabular}{lccc}
\toprule
Difficulty & Max.\ deformation  & Amplitude change & Min.\ coherence \\
           & (rad)             & (fraction)       &                 \\
\midrule
Easy   & 0.35--0.70 & 0.03--0.09 & 0.93--0.98 \\
Medium & 0.65--1.15 & 0.06--0.15 & 0.83--0.94 \\
Hard   & 1.00--1.55 & 0.10--0.22 & 0.70--0.86 \\
\bottomrule
\end{tabular}
\end{table}

Thus, an epoch pair includes both shared scattering structure and controlled
temporal changes rather than a global phase rotation. The reference
differential phase is computed from $x_2x_1^{*}$ using the convention
established in Section~\ref{subsec:differential_phase_task}.

The dataset contains 3000 training, 500 validation, and 500 sealed pairs,
generated from 188, 32, and 32 parent scenes, respectively. The assignment of
parent scenes is completed before rendering and patch extraction, so the three
splits share neither parent scenes nor overlapping crops. The stored files
contain clean complex references, differential-phase labels, reliability
weights, validity masks, crop coordinates, and scene metadata. Sampling masks
and noisy measurements are generated online; therefore, the same clean pair can
be evaluated under different acquisition budgets without storing a mask-specific
image target.

\subsubsection{Sentinel-1 external evaluation}

\begin{table*}[t]
\centering
\caption{Sentinel-1 scenes used for cross-domain evaluation.}
\label{tab:sentinel_scenes}
\scriptsize
\setlength{\tabcolsep}{3.2pt}
\begin{tabular}{cllcccrr}
\toprule
Scene & Surface type & Location & Satellite & Date pair & Rel. orbit & Patches & Mean coherence \\
\midrule
S1 & Mixed built-up coastal plain & Chiba--Ibaraki, Japan & 1B & 2021-12-07/2021-12-19 & 141 & 30 & 0.380 \\
S2 & Semi-arid bare land & Central Myanmar, west of Bagan & 1A & 2024-03-10/2024-03-22 & 4 & 30 & 0.314 \\
S3 & Urban-industrial coastal plain & Nagoya--Nobi, Japan & 1B & 2021-12-05/2021-12-17 & 112 & 30 & 0.494 \\
\bottomrule
\end{tabular}
\end{table*}

For cross-domain evaluation, we use three public Sentinel-1 C-band Level-1 IW
SLC repeat-pass pairs, which preserve complex amplitude and phase
\cite{Torres2012,Registry2026}.  All pairs use VV polarization, the IW1
subswath, a 12-day same-track interval, and no external-domain training.  The
scenes were fixed before inference and comprise a mixed built-up coastal
plain, semi-arid bare land, and a flat urban-industrial coastal plain; their
acquisition details and measured reference coherence are listed in Table~\ref{tab:sentinel_scenes}.

For each pair, integer-pixel translation is estimated using log-amplitude correlation, followed by removal of the dominant phase ramp and global phase offset. We retain 30 non-overlapping $256\times256$ patch pairs from each of Kanto, Bagan, and Nagoya, yielding 90 pairs in total. No patch is selected based on its mean coherence, and continuous coherence weighting is used for evaluation.

The registered SLC patches serve as complex references, whereas undersampled
observations are generated by the same frozen phase-history operator used for
the simulated dataset.  In particular, SLC image pixels are not deleted and then
reinterpreted as missing aperture samples.  This design isolates whether the learned prior transfers to real-SAR scene statistics, including speckle, terrain-induced phase, and natural coherence structure, rather than to a different acquisition geometry.  Because deformation ground truth is
unavailable, the experiment measures complex-reconstruction and wrapped
differential-phase consistency, not absolute displacement accuracy.

\subsection{Sampling and Observation Protocol}
\label{subsec:sampling_protocol}

All methods use the matrix-free phase-history operator introduced in
Section~\ref{subsec:observation_model}. It evaluates the centered
two-dimensional Fourier model at geometry-derived non-Cartesian spatial
frequencies with a matched interpolation/scatter pair; no dense sensing matrix
is constructed. From $N_a=256$ candidate aperture positions, we use exact-count
budgets $K\in\{80,104,128\}$, corresponding to sampling rates of 31.25\%,
40.625\%, and 50\%, referred to below as 30\%, 40\%, and 50\%. The three
approximately equally spaced budgets represent a severe undersampling
condition, an intermediate condition, and a half-aperture condition while
keeping the image grid, bandwidth, and operator unchanged.

The maximum-gap bound of \eqref{eq:feasible_mask_set} is fixed before any
experiment as
\begin{equation}
    g(K)=\max_{m\in\{\mathrm{Uni},\mathrm{Poi},\mathrm{Sta}\}}
    \operatorname{gap}(m,K)+2,
    \label{eq:gap_bound_rule}
\end{equation}
which gives $g=7$, $5$, and $4$ for $K=80$, $104$, and $128$. The margin of two
samples keeps every conventional pattern family inside the feasible set, so
that all methods search within the same admissible region. The bound is an
upper limit rather than a target. The realized gaps of the selected masks are
3, 3, and 2.

\begin{figure}[H]
    \centering
    \includegraphics[width=\columnwidth]%
    {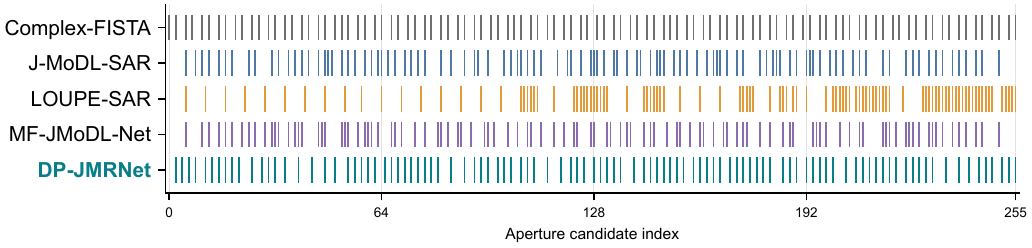}
    \caption{Aperture sampling patterns at the 40\% budget. J-MoDL-SAR,
    LOUPE-SAR, and MF-JMoDL-Net optimize their patterns during training,
    whereas Complex-FISTA and DP-JMRNet use fixed patterns.}
    \label{fig:k104_masks}
\end{figure}

We compare DP-JMRNet with four representative baselines: Complex-FISTA
\cite{Beck2009}, J-MoDL-SAR \cite{Aggarwal2020}, LOUPE-SAR \cite{Bahadir2020},
and MF-JMoDL-Net \cite{Wu2024}. Within each system, both epochs use the same
aperture mask. Complex-FISTA uses a fixed uniform mask, whereas J-MoDL-SAR,
LOUPE-SAR, and MF-JMoDL-Net learn their sampling patterns jointly with their
reconstructors. DP-JMRNet uses a fixed standard pattern, so the reported
margins are not attributable to sampling design. Figure~\ref{fig:k104_masks}
shows the resulting patterns at the intermediate budget.

The nominal evaluation SNR is 20~dB, which is the midpoint of the 15--25~dB
training range and therefore avoids choosing either a low-SNR failure case or a
high-SNR saturation case as the main operating point. Robustness is evaluated
separately below at 10, 15, 20, 25, and 30~dB. Noise is circular complex
Gaussian and is independent between epochs. The remaining system parameters are
reported in Appendix~\ref{app:experimental_details}.

\subsection{Evaluation Metrics}
\label{subsec:evaluation_metrics}

Let $a_{t}[p]=|x_{t}[p]|$ and $\hat a_{t}[p]=|\hat x_{t}[p]|$, and let
$e_{\Delta\phi}[p]$ and $e_{t}[p]$ denote the wrapped differential-phase and
single-epoch phase errors defined in \eqref{eq:differential_phase_error} and
\eqref{eq:single_epoch_phase_error}. The three primary metrics are the
reliability-weighted circular differential-phase RMSE (W-Diff), amplitude NMSE,
and complex NMSE:
\begin{align}
\mathrm{W\text{-}Diff}
&=\left(\frac{\sum_{p}w_{p}e_{\Delta\phi}^{2}[p]}
{\sum_{p}w_{p}+\varepsilon}\right)^{1/2},
\label{eq:metric_wdiff}\\
\mathrm{NMSE}_{\mathrm{amp}}
&=\frac{\sum_{t=1}^{2}\sum_{p}
\bigl(\hat a_{t}[p]-a_{t}[p]\bigr)^{2}}
{\sum_{t=1}^{2}\sum_{p}a_{t}^{2}[p]+\varepsilon},
\label{eq:metric_amp_nmse}\\
\mathrm{NMSE}_{\mathrm{cplx}}
&=\frac{\sum_{t=1}^{2}
\lVert\hat x_{t}-x_{t}\rVert_{2}^{2}}
{\sum_{t=1}^{2}\lVert x_{t}\rVert_{2}^{2}+\varepsilon}.
\label{eq:metric_complex_nmse}
\end{align}
W-Diff coincides with the differential-phase loss
\eqref{eq:differential_phase_loss} evaluated on held-out data, and is reported
in radians and degrees; lower values are better. Both NMSE measures are also
reported as $\mathrm{NMSE}_{\mathrm{dB}}=10\log_{10}(\mathrm{NMSE})$, for which
more negative values are better. W-Diff measures the final bitemporal phase
task, whereas the two NMSE metrics prevent a phase gain obtained by sacrificing
amplitude or overall complex reconstruction fidelity.

Table~\ref{tab:dpjmrnet_complete_metrics} also reports several secondary
metrics. Hard-Diff evaluates the differential-phase error over the frozen
hard-validity set, while W-Raw measures the reliability-weighted single-epoch
phase error. Amplitude fidelity is assessed using PSNR, SSIM \cite{Wang2004},
and Pearson correlation, with the reported values averaged over the two epochs.
PSLR, ISLR, and an AASR proxy are computed from the center-point response of
the frozen SAR operator using a fixed $3\times3$ mainlobe. These quantities
describe the sidelobe and azimuth-ambiguity behavior of the sampling pattern
\cite{Wu2024}. They are not used for checkpoint selection, and
Section~\ref{sec:results} shows that they do not predict differential-phase
performance across pattern families.

\subsection{Training Protocol}
\label{subsec:training_protocol}

DP-JMRNet uses five unfolding stages and 16 hidden channels. Training uses
AdamW \cite{Loshchilov2019} with a learning rate of $3\times10^{-4}$, weight
decay of $10^{-6}$, batch size 8, and gradient-norm clipping at 2.0. The loss
terms are those introduced in Section~\ref{sec:proposed_method}, and their
numerical weights are given in Appendix~\ref{app:experimental_details}. Detailed stopping rules, comparison-method settings, environment,
and timing scope are also provided in
Appendix~\ref{app:experimental_details}.

\section{Experimental Results and Discussion}
\label{sec:results}

This section evaluates DP-JMRNet along two complementary axes.  We first
characterize the overall performance of DP-JMRNet as the aperture budget
varies, which establishes whether the differential-phase gain persists across
sampling rates rather than arising at one convenient operating point.  We
then compare complete acquisition--reconstruction systems.  Accordingly, each system is run with its own sampling mask and matched reconstructor, and both
epochs share that mask within a system.

The comparison set spans the design choices that the proposed method
addresses.  Complex-FISTA \cite{Beck2009} is a classical sparse-recovery
reference on a fixed uniform mask, and therefore isolates the benefit of
learning either component.  J-MoDL-SAR \cite{Aggarwal2020} and LOUPE-SAR
\cite{Bahadir2020} learn the sampling pattern jointly with an unrolled or
feed-forward reconstructor, and MF-JMoDL-Net \cite{Wu2024} adopts a
SAR-specific joint design, but all three are supervised in the magnitude
domain.  None optimizes the bitemporal differential phase, and none
constrains the two temporal tracks to interact symmetrically.  J-MoDL and LOUPE were originally developed for MRI; we adapted them to SAR by replacing their MRI forward models with the same complex-valued SAR observation operator used in our experiments. MF-JMoDL-Net is designed for SAR reconstruction, but the authors did not release the source code. Therefore, the experiments below use our reproduction of MF-JMoDL-Net based on the published method.

\subsection{Behavior Across Aperture Budgets}
\label{subsec:overall_quantitative}

Table~\ref{tab:dpjmrnet_complete_metrics} summarizes the complete validation
performance of DP-JMRNet.  W-Diff RMSE decreases monotonically with the
aperture budget, from $10.3355^{\circ}$ at 30\% sampling rates to $9.0221^{\circ}$ at
50\% sampling rates.  Amplitude and complex NMSE improve over the same range, from
$-5.0620$ to $-7.4919$~dB and from $-1.2503$ to $-2.6632$~dB, as do PSNR,
SSIM, and correlation.  The differential-phase gain is therefore obtained
without any degradation of magnitude or complex fidelity.  SSIM and correlation are low in absolute terms because these optical-imagery measures penalize speckle that a physically consistent reconstruction reproduces only statistically. They serve here as relative diagnostics. For convenience, arrows are appended to the metrics in the table, where $\uparrow$ indicates that higher values are better and $\downarrow$ indicates that lower values are better.

\begin{table}[!ht]
\caption{Complete Validation Results of DP-JMRNet Under the Three Aperture
Budgets.}
\label{tab:dpjmrnet_complete_metrics}
\centering
\footnotesize
\renewcommand{\arraystretch}{1.06}
\setlength{\tabcolsep}{3.3pt}
\begin{tabular}{@{}lrrr@{}}
\toprule
& \multicolumn{3}{c}{Sampling rate} \\
\cmidrule(l){2-4}
Metric & 30\% & 40\% & 50\% \\
\midrule
\multicolumn{4}{@{}l}{\textit{Differential-phase and reconstruction metrics}} \\
W-Diff RMSE (rad) $\downarrow$       & 0.1804  & 0.1754  & \textbf{0.1575} \\
W-Diff RMSE (deg) $\downarrow$       & 10.3355 & 10.0488 & \textbf{9.0221} \\
Hard-Diff RMSE (rad) $\downarrow$    & 0.1573  & 0.1475  & \textbf{0.1294} \\
W-Raw RMSE (rad) $\downarrow$        & 1.1768  & 1.0520  & \textbf{0.9196} \\
Amplitude NMSE $\downarrow$          & 0.3130  & 0.2643  & \textbf{0.1808} \\
Amplitude NMSE (dB) $\downarrow$     & $-5.0620$ & $-5.8047$ & $\mathbf{-7.4919}$ \\
Complex NMSE $\downarrow$            & 0.7500  & 0.6560  & \textbf{0.5421} \\
Complex NMSE (dB) $\downarrow$       & $-1.2503$ & $-1.8328$ & $\mathbf{-2.6632}$ \\
\midrule
\multicolumn{4}{@{}l}{\textit{Image-quality and acquisition diagnostics}} \\
PSNR (dB) $\uparrow$                 & 14.7320 & 15.4747 & \textbf{17.1624} \\
SSIM $\uparrow$                      & 0.2239  & 0.2695  & \textbf{0.3951} \\
Correlation $\uparrow$               & 0.2141  & 0.2563  & \textbf{0.3818} \\
PSLR (dB) $\downarrow$               & $\mathbf{-11.5652}$ & $-9.4146$ & $-11.2957$ \\
ISLR (dB) $\downarrow$               & 10.4189 & 9.3484 & \textbf{8.1884} \\
AASR proxy (dB) $\downarrow$         & $-37.5873$ & $-38.7466$ & $\mathbf{-39.8952}$ \\
\bottomrule
\end{tabular}
\end{table}

\begin{table*}[!t]
\caption{Full-System Validation Comparison Using the Sampling Pattern and
Matched Reconstructor Associated With Each Method.}
\label{tab:main_method_comparison}
\centering
\footnotesize
\renewcommand{\arraystretch}{1.08}
\setlength{\tabcolsep}{3.7pt}
\begin{tabular}{@{}lrrrrrrrrr@{}}
\toprule
& \multicolumn{3}{c}{30\% sampling rate}
& \multicolumn{3}{c}{40\% sampling rate}
& \multicolumn{3}{c}{50\% sampling rate} \\
\cmidrule(lr){2-4}\cmidrule(lr){5-7}\cmidrule(l){8-10}
Method
& W-Diff & Amp. & Complex
& W-Diff & Amp. & Complex
& W-Diff & Amp. & Complex \\
& (deg) & (dB) & (dB)
& (deg) & (dB) & (dB)
& (deg) & (dB) & (dB) \\
\midrule
Complex-FISTA
& 35.2916 & $-1.7962$ & $-0.6231$
& 32.9246 & $-2.3797$ & $-1.0783$
& 30.5585 & $-2.9213$ & $-1.5455$ \\
J-MoDL-SAR
& 21.6838 & $-4.4497$ & $\mathbf{-1.5088}$
& 21.2877 & $-5.3458$ & $\mathbf{-2.0720}$
& 21.6812 & $-6.2819$ & $\mathbf{-2.6997}$ \\
LOUPE-SAR
& 12.1637 & $\mathbf{-6.9225}$ & 1.5796
& 19.1496 & $-4.9341$ & $-1.3361$
& 16.1613 & $-4.4418$ & $-0.8438$ \\
MF-JMoDL-Net
& 19.8557 & $-2.4126$ & $-0.9840$
& 19.2690 & $-5.0172$ & $-1.4592$
& 18.5188 & $-5.2568$ & $-1.9584$ \\
DP-JMRNet
& \textbf{10.3355} & $-5.0620$ & $-1.2503$
& \textbf{10.0488} & $\mathbf{-5.8047}$ & $-1.8328$
& \textbf{9.0221} & $\mathbf{-7.4919}$ & $-2.6632$ \\
\bottomrule
\end{tabular}

\vspace{2pt}
\begin{minipage}{0.98\textwidth}
\end{minipage}
\end{table*}
The acquisition diagnostics behave differently.  ISLR and the AASR proxy
improve monotonically with the budget, but PSLR is best at 30\% sampling rate
($-11.5652$~dB), which is the budget with the worst differential-phase
accuracy.  A point-response statistic summarizes the impulse response of the
acquisition geometry alone and carries no information about how
reconstruction error distributes between the two epochs, which is what the
differential phase depends on.  This dissociation motivates the systematic analysis in
Section~\ref{subsec:pattern_limits}, which shows that point-response
statistics do not predict differential-phase performance across aperture
pattern families.

\subsection{Full-System Performance and Robustness}
\label{subsec:fullcomparison}

Table~\ref{tab:main_method_comparison} reports the primary comparison.
DP-JMRNet obtains the lowest W-Diff RMSE at every aperture budget.  Relative
to the best baseline whose checkpoint satisfied the prespecified quality
criterion, the reduction is 47.9\%, 47.5\%, and 51.3\% at 30\%, 40\%, and 50\% sampling rate; relative to J-MoDL-SAR, the closest architectural analogue, it is
52.3\%, 52.8\%, and 58.4\%.

\begin{figure*}[t]
    \centering
    \includegraphics[width=\textwidth]{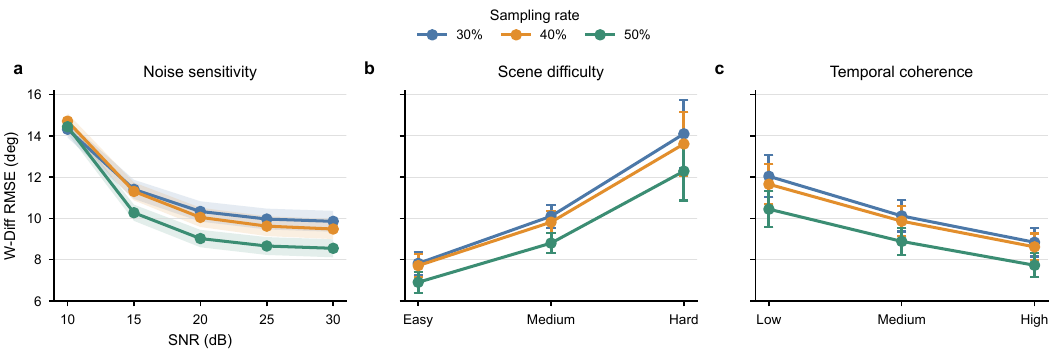}
    \caption{Robustness of DP-JMRNet across operating conditions and aperture
    budgets.}
    \label{fig:robustness}
\end{figure*}

Two baseline behaviors support the motivation of this work.  The
differential-phase accuracy of J-MoDL-SAR is essentially flat across budgets
($21.68^{\circ}$, $21.29^{\circ}$, $21.68^{\circ}$) although its amplitude
NMSE improves by 1.83~dB over the same range: additional measurements improve
what its objective supervises while leaving the interferometric quantity
unchanged.  More directly, the LOUPE-SAR checkpoint at 30\% sampling rate attains the
lowest amplitude NMSE in the table ($-6.9225$~dB) while producing a
\emph{positive} complex NMSE in dB ($+1.5796$~dB), meaning that the complex
error energy exceeds that of the reference and the phase content has been
destroyed while the magnitude envelope is preserved.  That checkpoint,
together with the one at 50\% sampling rate, failed the prespecified quality criterion
and is retained for transparency; magnitude-domain metrics alone would have
ranked it first.

J-MoDL-SAR retains a small complex NMSE advantage, narrowing from 0.26~dB at
30\% sampling rate to 0.04~dB at 50\% sampling rate, where the two systems are effectively tied.
Complex-FISTA is uniformly the weakest, as expected for a system with neither
a learned aperture pattern nor a learned prior.  Overall, DP-JMRNet is best
on W-Diff at all three budgets and best on amplitude NMSE at two of three,
conceding a fraction of a decibel on complex NMSE to a system whose
differential-phase error is more than twice as large. DP-JMRNet achieves a substantial improvement in differential-phase accuracy while preserving comparable performance in amplitude and complex reconstruction quality.

Beyond the nominal operating condition, we further examine whether the
differential-phase performance of DP-JMRNet remains stable under variations in
measurement noise, scene difficulty, and temporal coherence. Specifically, we evaluate W-Diff across SNRs from 10 to 30~dB, three levels of scene difficulty and temporal coherence, and all three aperture budgets of 30

As shown in Fig.~\ref{fig:robustness}(a), W-Diff decreases consistently as
the SNR increases from 10 to 30~dB, while the relative behavior across aperture
budgets remains stable. Similar trends are observed under varying scene
difficulty and temporal coherence in Figs.~\ref{fig:robustness}(b) and (c):
harder scenes and lower coherence increase reconstruction difficulty, but do
not alter the overall performance ordering. These results indicate that the
phase-preserving behavior of DP-JMRNet is not confined to the nominal 20~dB
setting or to a particular scene regime.


\subsection{Qualitative Reconstruction Results}
\label{subsec:qualitative_results}

\begin{figure*}[t]
\centering
\includegraphics[width=\textwidth]{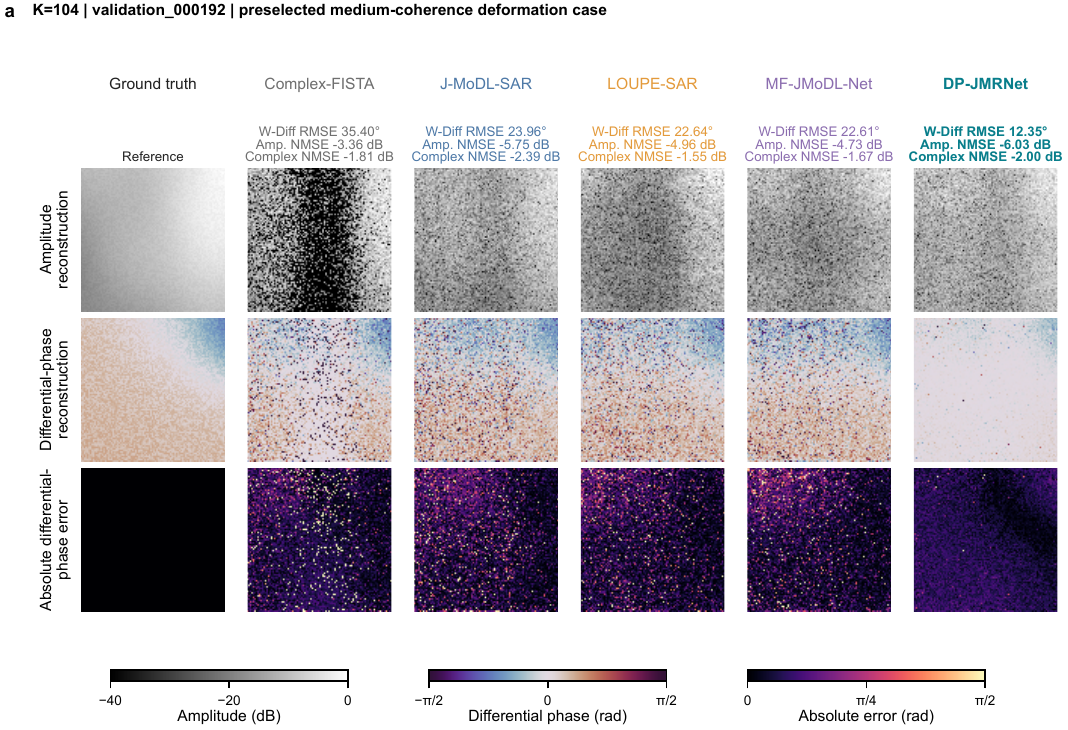}
\caption{Qualitative full-system comparison for a representative simulated
case at 40\% sampling rate.}
\label{fig:k104_qualitative}
\end{figure*}

Fig.~\ref{fig:k104_qualitative} presents a representative medium-coherence
deformation case at 40\% sampling rate. The columns correspond to the ground truth and the competing reconstruction methods, while the three rows show the amplitude reconstruction, differential-phase reconstruction, and absolute differential-phase error, respectively. Lower W-Diff RMSE and amplitude/complex NMSE indicate better reconstruction quality. Complex-FISTA exhibits pronounced amplitude
artifacts and spatially scattered differential-phase errors.  The learned
baselines improve the amplitude reconstruction, but residual phase outliers
remain visible, particularly in less reliable regions.

DP-JMRNet produces a markedly cleaner differential-phase field, with substantially fewer severe phase outliers and a more spatially localized error distribution than the competing methods. Meanwhile, it preserves the large-scale amplitude structure and achieves the best amplitude NMSE of $-6.03$~dB in this case. Its W-Diff RMSE is reduced to $12.35^{\circ}$, whereas J-MoDL-SAR attains a slightly lower complex NMSE. This contrast further indicates that improved complex-domain fidelity does not necessarily translate into better differential-phase preservation, and highlights the phase-preserving behavior of DP-JMRNet observed in the aggregate results.



\subsection{Ablation and Mechanism Analysis}
\label{subsec:ablation_mechanism}

\begin{figure}[h]
    \centering
    \includegraphics[width=\columnwidth]{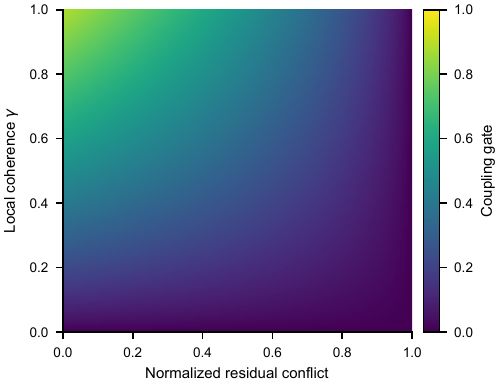}
    \caption{Continuous response of the learned selective gate.}
    \label{fig:gate_response}
\end{figure}

The selective gate is designed to strengthen inter-epoch information exchange only when the two epochs are locally reliable and mutually consistent. To examine what the gate has learned, we fix the normalized mean residual at 0.5 and evaluate the final-stage gate over different levels of local coherence and normalized residual conflict. As shown in Fig.~\ref{fig:gate_response}, the gate takes its largest values when coherence is high and residual conflict is low, indicating stronger information sharing between the two epochs. As coherence decreases or residual conflict increases, the gate value gradually decreases, thereby suppressing cross-epoch interaction when the two epochs become unreliable or inconsistent. Rather than making a hard share-or-isolate decision, the gate continuously adjusts the interaction strength and thus reduces cross-epoch contamination in difficult regions.

\begin{figure}[h]
    \centering
    \includegraphics[width=\columnwidth]{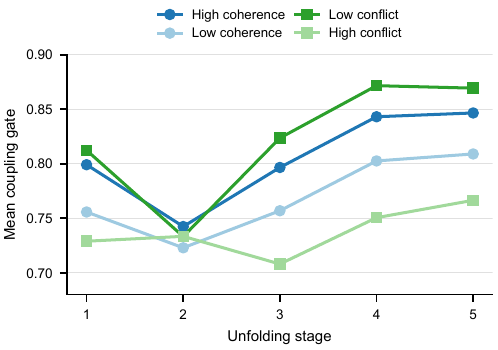}
    \caption{Stage-wise selective-sharing behavior.}
    \label{fig:stagewise_gate}
\end{figure}

We further examine whether this behavior persists throughout the unfolding
process. Figure~\ref{fig:stagewise_gate} reports the mean gate values across
the five stages at 40\% sampling, computed over eight audited validation cases
and stratified by local coherence and residual conflict. High-coherence regions
generally receive stronger coupling than low-coherence regions, whereas
high-conflict regions are suppressed relative to low-conflict regions. The
temporary reduction in the second stage indicates that the interaction is
stage adaptive rather than monotonically increased by construction.

We next intervene on the trained DP-JMRNet gate without changing any learned
parameters. Specifically, setting $g=0$ disables feature exchange, setting
$g=1$ enforces unconditional exchange, and the full model retains its
spatially varying continuous gate. The three settings are evaluated at 40\%
sampling using the same trained checkpoint, with lower values indicating
better performance for all metrics reported in Table~\ref{tab:gate_ablation}.
Disabling exchange increases W-Diff RMSE from $10.0488^{\circ}$ to
$19.3409^{\circ}$. Unconditional exchange recovers most of the phase accuracy
but produces a higher complex NMSE than the full model. The continuous gate
achieves the lowest W-Diff and complex NMSE, whereas the always-on
intervention gives the lowest amplitude NMSE.

Beyond differential phase, the interaction module also improves single-epoch
amplitude reconstruction: disabling exchange degrades the amplitude NMSE from
$-5.78$~dB to $-3.98$~dB, a gain of 1.80~dB. For reference, the amplitude
margin of DP-JMRNet over the best baseline is only 0.79~dB. This gain cannot
be attributed to a common-mode phase effect. Sharing information across epochs
can correlate the two phase errors, but it cannot reduce amplitude error by
that mechanism. The 1.80~dB improvement therefore indicates that the module
extracts complementary scene information rather than merely redistributing
phase error.

\begin{table}[h]
    \caption{Gate intervention results at 40\% sampling.}
    \label{tab:gate_ablation}
    \centering
    \footnotesize
    \renewcommand{\arraystretch}{1.08}
    \setlength{\tabcolsep}{3.2pt}
    \begin{tabular}{@{}lrrr@{}}
        \toprule
        Gate setting & W-Diff & Amp. & Complex \\
        & (deg) & NMSE & NMSE \\
        \midrule
        Exchange off   & 19.3409 & 0.4004 & 0.6579 \\
        Always on      & 10.2411 & \textbf{0.2458} & 0.6765 \\
        Selective gate & \textbf{10.0488} & 0.2643 & \textbf{0.6560} \\
        \bottomrule
    \end{tabular}
\end{table}

\begin{figure}[h]
    \centering
    \includegraphics[width=\columnwidth]
    {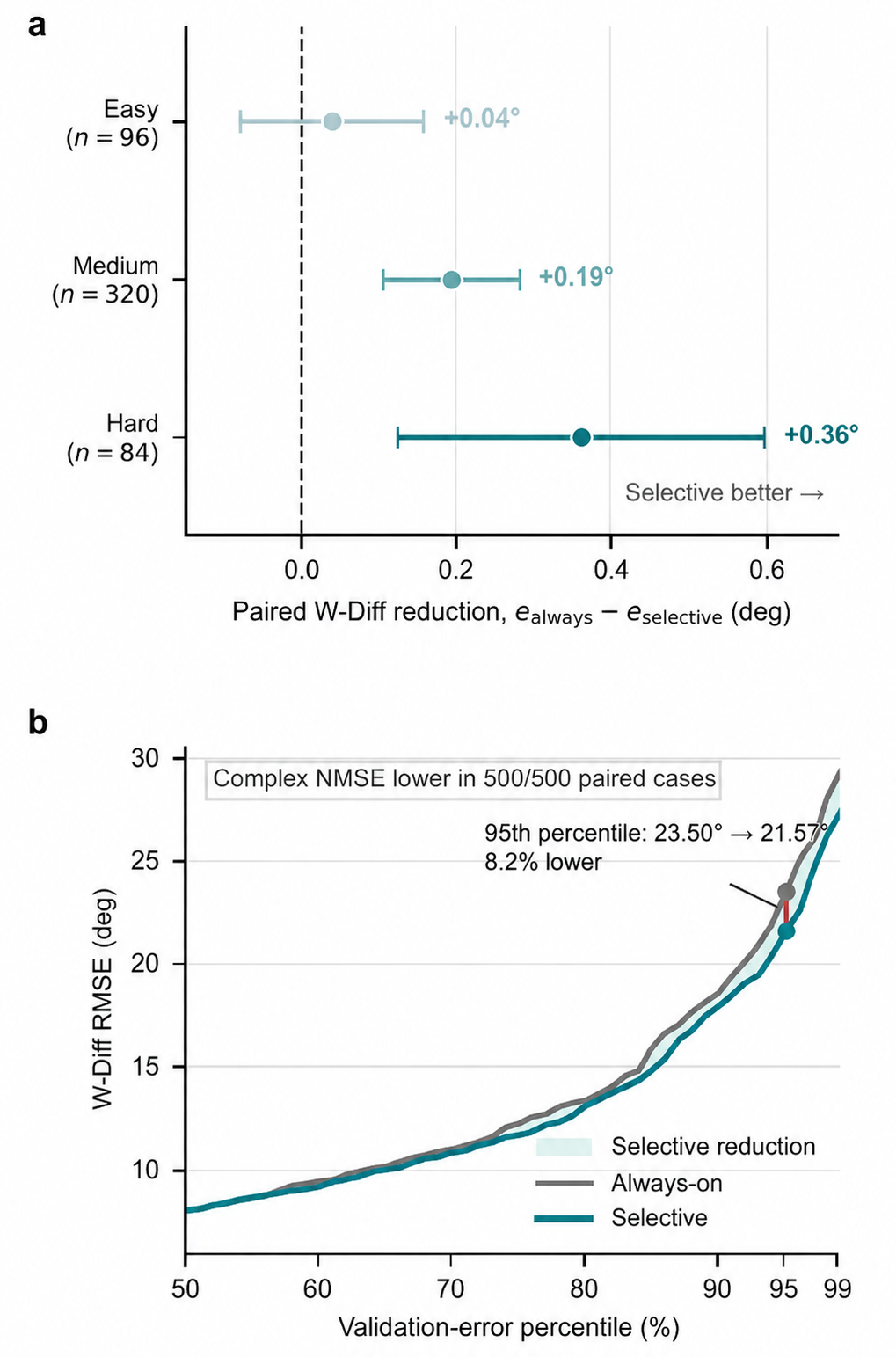}
    \caption{Robustness of selective interaction.}
    \label{fig:selective_necessity}
\end{figure}

The advantage of selective interaction becomes more pronounced as
reconstruction difficulty increases, as shown in
Fig.~\ref{fig:selective_necessity}. The paired W-Diff reduction increases from
$0.04^{\circ}$ for easy cases to $0.36^{\circ}$ for hard cases, while the
95th-percentile error decreases from $23.50^{\circ}$ to $21.57^{\circ}$. These
results indicate that selective interaction primarily improves robustness in
difficult and high-error cases, where unconditional feature exchange is more
likely to introduce adverse cross-temporal information.

\begin{figure}[t]
    \centering
    \includegraphics[width=\columnwidth]{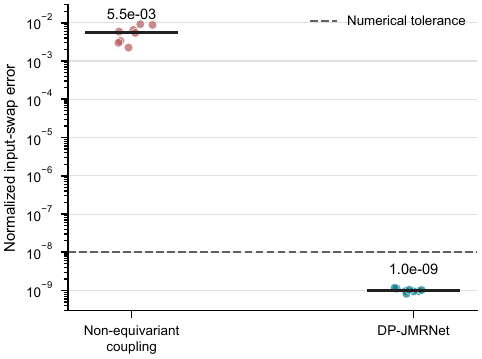}
    \caption{Normalized input-swap error.}
    \label{fig:exchange_equivariance}
\end{figure}

Finally, we assess the exchange-equivariant property directly by swapping the
two inputs and measuring the normalized output discrepancy. At 40\% sampling,
Fig.~\ref{fig:exchange_equivariance} shows the errors for eight audited cases;
each point denotes one case and the horizontal segment denotes the arithmetic
mean. The comparator is a separately trained non-equivariant coupled
architecture used only as a structural diagnostic. Its mean swap error is
$5.5\times10^{-3}$, whereas DP-JMRNet achieves a mean error of approximately
$1.0\times10^{-9}$ and a maximum error of $1.2\times10^{-9}$, both below the
predefined $10^{-8}$ numerical tolerance. DP-JMRNet therefore satisfies
exchange equivariance to floating-point precision.

\begin{figure*}[t]
    \centering
    \includegraphics[width=\textwidth]{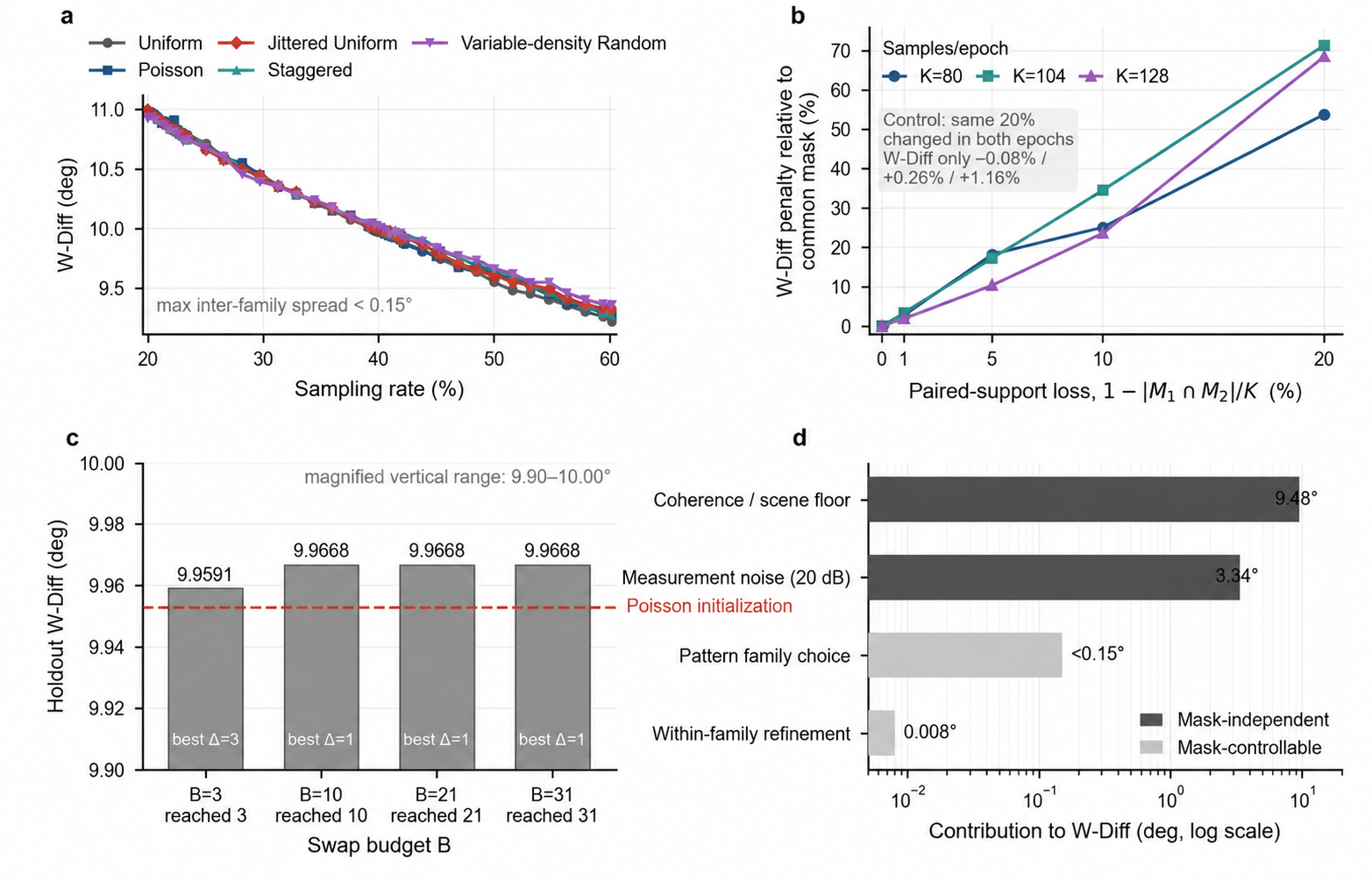}
    \caption{Analysis of the role and limitations of aperture-pattern design.}
    \label{fig:sampling_design_analysis}
\end{figure*}

To determine how much of the differential-phase performance can be attributed
to acquisition design, we isolated three aspects of aperture pattern design
while keeping the reconstructor fixed. The results are shown in 
Fig.~\ref{fig:sampling_design_analysis}.

\subsection{Role and Limits of Aperture Pattern Design}
\label{subsec:pattern_limits}

The first experiment measures the effect of the pattern family. Five families
were compared: Uniform, Poisson, Staggered, Jittered Uniform, and
Variable-density Random. Each was evaluated at 41 sampling rates between 20\%
and 60\%, and every pattern satisfied the prescribed sample count and the gap
constraint of \eqref{eq:feasible_mask_set}.
Fig.~\ref{fig:sampling_design_analysis}(a) shows that the five curves are almost
indistinguishable across the entire range, with a maximum spread of
$0.15^{\circ}$. For comparison, raising the sampling rate from 20\% to 60\%
reduces W-Diff by about $1.8^{\circ}$. Within the feasible set, therefore,
which family is used has little effect on the differential phase.

The second experiment tests whether the two epochs must share the same
sampling mask. Starting from a common mask, we replaced a fraction of
the second-epoch aperture positions so that the two epochs no longer sampled
identical locations, giving paired-support losses of 0\%, 1\%, 5\%, 10\%, and
20\%. As a control, the same fraction of positions was replaced in both
epochs, which changes the pattern by the same amount but keeps the two
supports identical. Fig.~\ref{fig:sampling_design_analysis}(b) shows a clear dose
response: losing 20\% of the paired support increases W-Diff by 53.8\% to
71.4\% across the three budgets, whereas the control changes it by only
$-0.08\%$ to $+1.16\%$. What degrades the differential phase is thus the
mismatch between epochs, not the perturbation of the pattern itself.

The third experiment examines whether a task-driven search within the feasible
set improves on a standard pattern. Starting from the same Poisson mask, a
cumulative remove--add search was run at 40\% sampling under four deviation
budgets, $B\in\{3,10,21,31\}$, where $B$ limits how far the search may move
from the initialization. Each budget received exactly 256 candidate
evaluations on 24 search scenes, and the selected masks were then evaluated on
128 disjoint holdout scenes. As shown in Fig.~\ref{fig:sampling_design_analysis}(c),
every search chain reached its allowed radius, so none was limited by
insufficient exploration. Nevertheless, all four returned masks are marginally
worse than the initialization on the holdout scenes, and the best solutions
remain within one to three replacements of it. Optimizing the aperture pattern for differential phase is therefore
ineffective: once the feasibility constraints are met, no pattern in the
feasible set improves on a standard one.

Fig.~\ref{fig:sampling_design_analysis}(d) places these effects on a common scale. Scene
coherence and measurement noise contribute $9.48^{\circ}$ and $3.34^{\circ}$
to W-Diff, whereas the family choice contributes less than $0.15^{\circ}$ and
within-family refinement about $0.008^{\circ}$. What aperture design can
control is two to three orders of magnitude smaller than what it cannot.

These observations share one explanation. Since both epochs are observed
through the same aperture support, an error introduced by the pattern appears
in both reconstructions and cancels in the conjugate product of
\eqref{eq:differential_phase_error}. This is why the family choice and the
refined masks make so little difference, and why breaking the pairing, which
removes the cancellation, is so damaging. The overall  conclusion is that
repeating the same aperture positions across epochs matters far more than
choosing which positions to use.

\subsection{Cross-Domain Evaluation on Sentinel-1 Data}
\label{subsec:sentinel_external}

Table~\ref{tab:sentinel_external} reports the equal-scene means over the three
fixed Sentinel-1 scenes, with 30 patch pairs per scene.  DP-JMRNet achieves
the lowest W-Diff at all three sampling rates.  Relative to Complex-FISTA,
LOUPE-SAR, and MF-JMoDL-Net, its absolute reductions range from
4.23$^{\circ}$ to 14.08$^{\circ}$.  The margin over the stronger J-MoDL-SAR
baseline is smaller but remains consistent in the pooled means: DP-JMRNet
reduces W-Diff by 2.108$^{\circ}$, 0.904$^{\circ}$, and 1.580$^{\circ}$ at
30\%, 40\%, and 50\% sampling, respectively.

\begin{table*}[t]
\centering
\caption{Cross-domain reconstruction results on three Sentinel-1 scenes}
\label{tab:sentinel_external}
\footnotesize
\renewcommand{\arraystretch}{1.08}
\setlength{\tabcolsep}{3.7pt}
\begin{tabular}{@{}lrrrrrrrrr@{}}
\toprule
& \multicolumn{3}{c}{30\% sampling}
& \multicolumn{3}{c}{40\% sampling}
& \multicolumn{3}{c}{50\% sampling} \\
\cmidrule(lr){2-4}\cmidrule(lr){5-7}\cmidrule(l){8-10}
Method
& W-Diff & Amp. & Complex
& W-Diff & Amp. & Complex
& W-Diff & Amp. & Complex \\
& (deg) & (dB) & (dB)
& (deg) & (dB) & (dB)
& (deg) & (dB) & (dB) \\
\midrule
Complex-FISTA
& 78.7527 & $-2.3476$ & $-1.3960$
& 74.2875 & $-3.1834$ & $-2.1452$
& 69.4830 & $-3.8752$ & $-2.8169$ \\
J-MoDL-SAR
& 76.0918 & $\mathbf{-3.8314}$ & $\mathbf{-1.7009}$
& 70.6079 & $\mathbf{-4.9608}$ & $\mathbf{-2.4753}$
& 65.8025 & $\mathbf{-5.9358}$ & $\mathbf{-3.2107}$ \\
LOUPE-SAR
& 81.2619 & $-3.7370$ & 1.4537
& 76.9477 & $-4.0552$ & $-1.4423$
& 78.3055 & $-3.2855$ & $-0.9173$ \\
MF-JMoDL-Net
& 78.2155 & $-2.2043$ & $-1.1244$
& 75.4875 & $-4.1066$ & $-1.5959$
& 71.3029 & $-4.5320$ & $-2.1351$ \\
DP-JMRNet
& $\mathbf{73.9837}$ & $-3.0884$ & $-1.2368$
& $\mathbf{69.7037}$ & $-3.8103$ & $-1.9145$
& $\mathbf{64.2220}$ & $-4.5742$ & $-2.5879$ \\
\bottomrule
\end{tabular}
\end{table*}

The metrics are consistent with the simulated results.
DP-JMRNet leads on W-Diff at every sampling rate while J-MoDL-SAR retains a small advantage in amplitude and complex NMSE, with the gap staying below
1.1~dB and 0.7~dB, respectively. The remaining baselines are behind DP-JMRNet on the differential phase by a wide margin and offer no compensating advantage elsewhere. The differential-phase preference learned on simulated data
therefore transfers to measured SAR scenes without any retraining, and it does so without a corresponding loss in image reconstruction quality.

Absolute W-Diff values are high across all methods, between $64^{\circ}$ and $81^{\circ}$, which reflects the low reference coherence of real scenes rather than a failure of any particular reconstructor. Under such conditions the differential phase is intrinsically noisy, and the consistent ranking
across scenes and sampling rates is more informative than the absolute level.
The result indicates that the proposed method remains usable in realistic
low coherence settings, where the interferometric task is hardest.


\subsection{Computational Characteristics and Limitations}
\label{subsec:cost_limitations}

The implementation and timing conditions are reported in Appendix. DP-JMRNet contains only 122,533 trainable parameters, approximately one third of the 370.5k parameters used by J-MoDL-SAR and MF-JMoDL-Net, and substantially fewer than the 31.39 million parameters of LOUPE-SAR. Its dedicated 40\%-sampling, batch-1 latency is 105.95~ms per sample,
whereas the full-validation measurements report amortized per-sample times of
20.3, 43.2, and 20.4~ms at 30\%, 40\%, and 50\% sampling
rates, respectively. Because these measurements use different batching and
timing scopes, they are reported as computational characteristics and are not
converted into a speed--quality comparison figure.

\section{Conclusions}
\label{sec:conclusion}

This paper presented DP-JMRNet, a physics-guided framework for complex
bitemporal SAR reconstruction under a differential-phase objective. The
reconstructor combines matrix-free SAR data consistency with complex-domain
deep unfolding, while exchange-equivariant selective interaction enables
information sharing between epochs without introducing an artificial
dependence on their input order.

Experiments at 30\%, 40\%, and 50\% sampling rates showed that DP-JMRNet
consistently achieved the lowest W-Diff RMSE, reducing it by 47.5\%--51.3\%, while
maintaining competitive amplitude and complex-image fidelity. The ablation
results further demonstrated that selective interaction is particularly
beneficial for difficult and high-error cases, and that it improves
single-epoch amplitude reconstruction by 1.80~dB, indicating genuine extraction
of complementary information. The input-swap audit verified exchange
equivariance to numerical precision. Cross-domain evaluation on three
Sentinel-1 scenes produced the lowest W-Diff at all three sampling rates,
indicating that the learned differential-phase preference transfers to measured
SAR data.

We also characterized the role of acquisition design. Sharing the same aperture
support across epochs is necessary for phase fidelity, since breaking it
degrades W-Diff by more than 50\%. Within the feasible set, however, neither
the choice of pattern family nor task-driven refinement changes the
differential phase appreciably, because pattern-induced errors are common mode
and cancel in the interferogram. Acquisition design therefore matters mainly
through its constraints rather than through fine-grained pattern optimization. This is consistent with an observation in \cite{Wu2024}, where learned sampling patterns
were found to approach uniform sampling as scene sparsity decreased. Our
results indicate that the benefit of pattern optimization depends not only on
the scene but also on the task: for differential-phase preservation, distinct
feasible patterns become nearly indistinguishable.

Despite these results, several limitations remain. The remaining degradation
under severe noise and low temporal coherence shows that reliable phase
recovery is still challenging in strongly decorrelated regions. Moreover, the
study focuses on differential-phase preservation at the complex-SAR
reconstruction level and does not directly evaluate deformation retrieval using
a complete InSAR processing chain. Future work will therefore integrate
DP-JMRNet with InSAR processing and assess whether the improved
differential-phase fidelity translates into more accurate line-of-sight
deformation estimates on measured datasets with reliable deformation
references.

\section*{Code Availability}

The code and data
of this study are publicly available at:
\url{https://github.com/JasonBao05/coherent-sar-unfolding}.

\appendix[Detailed Experimental Parameters]
\label{app:experimental_details}

Only parameters not stated in Section~\ref{sec:experimental_setup} are listed in
Tables~\ref{tab:appendix_acquisition}--\ref{tab:appendix_training}.

\begin{table}[H]
\centering
\caption{SAR Acquisition and Observation Parameters}
\label{tab:appendix_acquisition}
\footnotesize
\renewcommand{\arraystretch}{1.10}
\setlength{\tabcolsep}{5pt}
\begin{adjustbox}{max width=\columnwidth,center}
\begin{tabular}{cc}
\toprule
Parameter & Value \\
\midrule
Ground-plane spacing & $0.5\times0.5$ m \\
Center frequency & 10 GHz \\
Bandwidth & 400 MHz \\
Frequency samples & 256 \\
Aperture length & 4 m \\
Ground range & 100 m \\
Platform altitude & 100 m \\
Aperture chunk size & 32 \\
Maximum aperture gap at $K=80$ & 7 samples \\
Maximum aperture gap at $K=104$ & 5 samples \\
Maximum aperture gap at $K=128$ & 4 samples \\
\bottomrule
\end{tabular}
\end{adjustbox}
\end{table}

\begin{table}[h]
\centering
\caption{Implementation and Runtime Parameters}
\label{tab:appendix_runtime}
\footnotesize
\renewcommand{\arraystretch}{1.10}
\setlength{\tabcolsep}{5pt}
\begin{adjustbox}{max width=\columnwidth,center}
\begin{tabular}{cc}
\toprule
Parameter & Value \\
\midrule
Python version & 3.11.4 \\
PyTorch version & 2.7.1 \\
CUDA version & 12.8 \\
GPU & NVIDIA RTX 5070, 12 GB \\
Amortized time at $K=80$ & 20.3 ms/sample \\
Amortized time at $K=104$ & 43.2 ms/sample \\
Amortized time at $K=128$ & 20.4 ms/sample \\
Batch-1 latency at $K=104$ & 105.95 ms/sample \\
Latency measurement runs & 30 forward passes \\
Peak inference GPU memory & 76.0 MiB \\
\bottomrule
\end{tabular}
\end{adjustbox}
\end{table}

\begin{table}[h]
\centering
\caption{Model, Loss, and Training Termination Parameters}
\label{tab:appendix_training}
\footnotesize
\renewcommand{\arraystretch}{1.10}
\setlength{\tabcolsep}{5pt}
\begin{adjustbox}{max width=\columnwidth,center}
\begin{tabular}{cc}
\toprule
Parameter & Value \\
\midrule
Model coherence window & $7\times7$ \\
Sentinel reliability window & $9\times9$ \\
Complex loss weight & 1.0 \\
Amplitude loss weight & 0.35 \\
Single-epoch phase loss weight & 0.12 \\
Differential-phase loss weight & 1.25 \\
Gate smoothness weight & $10^{-4}$ \\
Anti-collapse weight & $10^{-3}$ \\
Coherence threshold & 0.70 \\
Mean-residual threshold & 0.30 \\
Conflict threshold & 0.30 \\
Minimum mean activation & 0.35 \\
Maximum training updates & 10\,000 \\
Minimum training updates & 1800 \\
Early-stopping patience & 300 updates \\
Baseline maximum updates & 800 \\
Baseline validation interval & 50 updates \\
Baseline patience & 100 updates \\
Baseline minimum updates & 100 \\
Baseline effective batch size & 4 \\
Complex-FISTA iterations & 40 \\
Complex-FISTA $\ell_1$ weight & 0.002 \\
\bottomrule
\end{tabular}
\end{adjustbox}
\end{table}

The maximum update values serve only as training caps. For all learned methods, the reported results are obtained from the checkpoint achieving the best validation performance under the corresponding early stopping criterion, rather than from the final training update.

\bibliographystyle{IEEEtran}
\bibliography{refs}

\end{document}